\documentclass[a4paper,12pt]{article}

\usepackage[T1]{fontenc} 
\usepackage{mathtools}
\usepackage{setspace}
\usepackage{slashed}
\usepackage{amsfonts}
\usepackage{amssymb}
\usepackage{graphicx}
\usepackage{caption}
\usepackage{subcaption}
\usepackage{array}
\usepackage{booktabs}
\usepackage{multirow}
\usepackage[normalem]{ulem}
\usepackage{color}
\usepackage{soul}
\usepackage{cite}
\usepackage{hyperref}

\newcommand{\bra}[1]{\left\langle #1 \right|}
\newcommand{\ket}[1]{\left| #1 \right\rangle}

\begin{document}

\hfill {\tt CERN-TH-2026-001}

\def\thefootnote{\fnsymbol{footnote}}

\begin{center}
\Large\bf\boldmath
\vspace*{2.cm} 
$\Lambda_b \to \Lambda$ Form Factors from Light-Cone Sum Rules with $\Lambda$-Baryon Distribution Amplitudes 
\end{center}
\vspace{1.cm}

\begin{center}
F. Mahmoudi$^{1,2,3}$\footnote{Electronic address: nazila@cern.ch}, 
D. Mishra$^{1}$\footnote{Electronic address: d.mishra@ip2i.in2p3.fr}\\
\vspace{1.cm}
{\sl $^1$Universit\'e Claude Bernard Lyon 1, CNRS/IN2P3, \\
Institut de Physique des 2 Infinis de Lyon, UMR 5822, F-69622, Villeurbanne, France}\\[0.4cm]
{\sl $^2$Theoretical Physics Department, CERN, CH-1211 Geneva 23, Switzerland}\\[0.4cm]
{\sl $^3$Institut Universitaire de France (IUF), 75005 Paris, France }\\[0.4cm]
\end{center}
\renewcommand{\thefootnote}{\arabic{footnote}}
\setcounter{footnote}{0}

\vspace{1.cm}
\begin{abstract}
\noindent

We compute the $\Lambda_b \to \Lambda$ transition form factors using light-cone sum rules based on $\Lambda$-baryon distribution amplitudes, including contributions up to twist--6. The analysis is performed for pseudoscalar, vector, and axial-vector $b\to s$ currents and for different choices of the $\Lambda_b$ interpolating current. A central aspect of the study is the explicit inclusion of the negative-parity $\Lambda_b^*$ state in the hadronic dispersion relation. Suitable linear combinations of invariant amplitudes are constructed to eliminate the $\Lambda_b^*$ pole contribution at the level of the sum rules. All three interpolating currents lead to numerically stable sum rules in the adopted working windows. The pseudoscalar current gives the closest agreement with the expected EOM and heavy quark symmetry relations at the level of the central values, while the present leading-order uncertainties do not allow a definitive discrimination among the three currents. The resulting helicity form factors are parametrised using a $z$-expansion and, for the pseudoscalar current, are combined with lattice-QCD tensor form factors to illustrate the low-$q^2$ differential branching fraction in $\Lambda_b \to \Lambda \ell^+ \ell^-$.

\end{abstract}

\thispagestyle{empty}

\clearpage

\section{Introduction} 
Rare decays of bottom hadrons induced by the flavour-changing neutral-current (FCNC) transition $b\to s \ell^+ \ell^-$ provide a sensitive probe of the Standard Model (SM) and of possible effects of physics beyond it. Being loop-suppressed in the SM, these processes are particularly sensitive to contributions from heavy degrees of freedom, while at the same time posing a significant theoretical challenge due to the presence of nonperturbative QCD dynamics. Over the past decade, extensive experimental studies of exclusive $b\to s \ell^+ \ell^-$ decays have been performed at the LHC, most notably by the LHCb collaboration, revealing several tensions between theoretical predictions and data in mesonic channels such as $B\to K^{(*)} \ell^+ \ell^-$ and $B_s\to \phi \ell^+ \ell^-$ (see~\cite{Hurth:2025vfx} for a recent analysis).

In this context, baryonic decays of the type $\Lambda_b \to \Lambda \ell^+ \ell^-$ offer a complementary laboratory for testing the flavour sector~\cite{Chen:2001zc,Aliev:2010uy,CDF:2011buy,Detmold:2012vy,LHCb:2013uqx,Liu:2015kaa,LHCb:2015tgy,Meinel:2016grj,Faustov:2017wbh,Bhattacharya:2019der,Meinel:2020owd,Bordone:2021bop,Blake:2022vfl,Amhis:2022vcd,Lu:2025gjt,Das:2020cpv,Ali:2025xkw,Das:2022xjg}. In contrast to mesonic modes, baryonic transitions provide access to a richer set of angular observables and helicity structures~\cite{Gutsche:2013pp,Boer:2014kda,Kumar:2015tnz,Blake:2017une,Descotes-Genon:2019dbw,Li:2022nim,Khan:2025rgc}, and a complementary new physics sensitivity (see e.g.~\cite{Hiller:2001zj,Aslam:2008hp,Wang:2008ni,Sahoo:2009zz,Azizi:2013eta,Azizi:2015hoa,Wang:2016cdw,Wang:2016dne,Alnahdi:2017ogx,Das:2018sms,Nasrullah:2018puc,Nasrullah:2018vky,Nasrullah:2020glp,Amhis:2020phx,Das:2022xjg,Biswas:2022fyb}). With the increasing precision of measurements~\cite{CDF:2011buy,LHCb:2013uqx,LHCb:2015tgy} in the baryonic sector, a reliable theoretical determination of the hadronic form factors governing the $\Lambda_b \to \Lambda$ transition has become essential.

The local form factors constitute an essential hadronic input for exclusive $b\to s \ell^+ \ell^-$ decays. In the low $q^2$ region, where the final-state hadron carries large energy, light-cone sum rules (LCSR) provide a well-established framework for computing the local form factors in a systematic expansion in terms of hadron distribution amplitudes (DAs). While LCSR methods have been extensively applied to heavy-to-light mesonic transitions, their extension to baryonic decays is technically more involved and remains comparatively less explored (see~\cite{Wang:2008sm,wang:2009hra,Aliev:2010uy,Khodjamirian:2011jp,Mannel:2011xg,Gan:2012tt,Wang:2015ndk,Huang:2024oik,Lu:2025gjt}).

In this work, we revisit the determination of the $\Lambda_b \to \Lambda$ transition form factors within the framework of LCSRs based on $\Lambda$ baryon DAs. A central issue in baryonic LCSR analyses is the sensitivity to the choice of interpolating current and the contamination from nearby excited states, in particular the negative-parity ($\Lambda_b^*$), whose mass lies close to the ground state. In the present analysis, we explicitly include both the $\Lambda_b(1/2^+)$ ground state and the negative-parity $\Lambda_b^*(1/2^-)$ excitation in the hadronic representation. We then construct suitable linear combinations of invariant amplitudes that eliminate the $\Lambda_b^*(1/2^-)$ pole contribution at the level of the sum rules, following and extending strategies previously developed for $\Lambda_b \to N$ transition~\cite{Khodjamirian:2011jp,Aliev:2018hyy}. 

Within this framework, we perform a systematic analysis of the pseudoscalar, axial-vector, and Ioffe interpolating currents for the $\Lambda_b$ baryon. All three choices lead to numerically stable sum rules in the adopted Borel and continuum-threshold windows. At the level of central values, the pseudoscalar current gives the closest agreement with the expected equation-of-motion (EOM) and heavy quark symmetry (HQS) relations, while the present leading-order uncertainties do not allow a definitive discrimination among the three currents. We include all three-quark $\Lambda$ DAs up to twist--$6$ and consider pseudoscalar, vector, and axial-vector $b\to s$ transition currents.

The resulting vector and axial-vector form factors are transformed into the helicity basis and parametrised using a truncated $z$-expansion, which also facilitates comparison with lattice-QCD results. As an illustrative phenomenological application, we combine the pseudoscalar current LCSR form factors with lattice-QCD tensor form factors and evaluate the differential branching fraction for $\Lambda_b \to \Lambda \ell^+ \ell^-$ in the low-$q^2$ region.

The paper is organised as follows. In Sec.~\ref{sec:corrl}, we introduce the correlation functions and their hadronic representations. Sec.~\ref{sec:lcope} describes the light-cone operator product expansion (OPE) and the construction of the dispersion relations. The resulting sum rules for the $\Lambda_b \to \Lambda$ form factors are presented in Sec.~\ref{sec:lcsr-ff}. Sec.~\ref{sec:result} contains the numerical analysis, the comparison of interpolating currents, and the phenomenological application to $\Lambda_b \to \Lambda \ell^+ \ell^-$. We summarise our findings in Sec.~\ref{sec:summary}. Technical details are collected in the appendices.

\section{Correlation function and hadronic representation}
\label{sec:corrl}
The starting point of the light-cone sum-rule analysis of the $\Lambda_b \to \Lambda$ transition form factors is the following vacuum-to-$\Lambda$ correlation function:
\begin{equation}
F_i^{(l)}(p,q)= i\!\int d^4x\, e^{iq\cdot x}\,
\langle 0|\,T\!\left\{\eta^{(l)}_{\Lambda_b}(0),\, j_i(x)\right\}|\Lambda(p)\rangle\,,
\label{eq:corrl}
\end{equation}
where $\eta^{(l)}_{\Lambda_b}$ interpolates the $\Lambda_b$ baryon, $j_i$ is the $b\to s$ weak transition current, and the momentum in the $\Lambda_b$ channel is $p_{\Lambda_b}=p-q$, which defines the variable of the hadronic dispersion relation.

The weak transition currents are chosen as:  
\begin{eqnarray}
    j_i(x) = \bar{s}(x)\, \Gamma_i\, b(x)\,,
\end{eqnarray}
with\footnote{
In this work, we do not consider the tensor weak current
$\bar{s}\,\sigma_{\mu\nu}(\gamma_5)\,b$.
Including tensor currents would require an independent set of correlation functions and tensor form factors, which are beyond the scope of the present analysis.
}
\begin{equation}
\Gamma_i \in \left\{ m_b\, i\gamma_5,\ \gamma_\mu,\ \gamma_\mu\gamma_5 \right\}.
\label{eq:trans_currents}
\end{equation} 
Here, we neglect $m_s$ in the overall normalisation of the pseudoscalar current.

A central ingredient of the LCSR construction is the choice of the interpolating current for the $\Lambda_b$ baryon. Different local currents carrying the same quantum numbers can couple with different strengths to the ground and excited states and may therefore lead to different numerical behavior of the resulting sum rules. In the present analysis, we consider three commonly used choices for the $\Lambda_b$ interpolating current: the pseudoscalar, axial-vector, and Ioffe-type currents~\cite{Wang:2008sm,Aliev:2010uy,Gan:2012tt,Aliev:2018hyy},  
\begin{eqnarray}
\eta^{({\cal P})} & = &
\epsilon^{abc}\left(u_a \,C\,\gamma_5\,d_b \right)b_c\,, \nonumber \\
\eta^{({\cal A})} & = &
\epsilon^{abc}\left(u_a\,C\,\gamma_5\gamma_\lambda\,d_b\right)\gamma^\lambda\,b_c\,, \nonumber \\
\eta^{({\cal I})} & = &
\epsilon^{abc}\left(u_a\,C\,\gamma_\lambda\,d_b\right)\gamma^\lambda\gamma_5\,b_c \,,
\label{eq:etas}
\end{eqnarray}
where $a,b,c$ are colour indices and $C$ denotes the charge-conjugation matrix. All three currents carry the quantum numbers of the $\Lambda_b$ baryon, but differ in their Dirac structures and coupling patterns to the ground and excited states.

The corresponding correlation functions $F_i^{(l)}$ are labeled by the choice of interpolating current, 
\begin{align}
F_i^{(\mathcal{P})} = \{F_{5}^{(\mathcal{P})},\, F_{\mu}^{(\mathcal{P})},\, F_{\mu 5}^{(\mathcal{P})}\}\,, \quad
F_i^{(\mathcal{A})} = \{F_{5}^{(\mathcal{A})},\, F_{\mu}^{(\mathcal{A})},\, F_{\mu 5}^{(\mathcal{A})}\}\,, \quad
F_i^{(\mathcal{I})} = \{F_{5}^{(\mathcal{I})},\, F_{\mu}^{(\mathcal{I})},\, F_{\mu 5}^{(\mathcal{I})}\}\,.
\end{align}
This classification allows for a systematic comparison of the resulting sum rules and their numerical behaviour for different choices of the interpolating current.

We now proceed to the hadronic representation of the correlation function by inserting a complete set of intermediate states with the quantum numbers of the $\Lambda_b$ baryon between the interpolating current $\eta^{(l)}_{\Lambda_b}$ and the weak transition current. As a result, the correlation function admits a dispersion representation in the variable $(p-q)^2$, whose lowest-lying contributions arise from the $\Lambda_b$ ground state and its negative-parity excitation, denoted by $\Lambda_b^*$.

The pole contribution of the $\Lambda_b$ ground state to the hadronic representations is determined by two matrix elements. The first one describes the coupling of the interpolating current to the $\Lambda_b$ ground state,
\begin{equation}
\langle 0 | \eta^{(l)}_{\Lambda_b} | \Lambda_b(p-q) \rangle
= m_{\Lambda_b}\, \lambda_{\Lambda_b}^{(l)}\, u_{\Lambda_b}(p-q)\,,
\label{eq:decayc}
\end{equation}
where $\lambda_{\Lambda_b}^{(l)}$ denotes the coupling associated with the interpolating current $\eta^{(l)}_{\Lambda_b}$, and $u_{\Lambda_b}(p-q)$ is the Dirac spinor of the $\Lambda_b$ baryon with momentum $p-q$.
The second matrix element encodes the weak transition between the intermediate $\Lambda_b$ state and the external $\Lambda$ baryon, $\langle \Lambda(p) | j_i | \Lambda_b(p-q) \rangle$, and is parametrised in terms of $\Lambda_b \to \Lambda$ form factors. We use the standard $\Lambda_b\to\Lambda$ form-factor conventions for the transition matrix elements below; an overall phase associated with the state conventions is immaterial for the resulting sum rules.  

For the pseudoscalar weak transition current, the hadronic matrix element is parameterised in terms of a single form factor as:  
\begin{equation}
\langle \Lambda(p) | m_b \,\bar{s}\, i\gamma_5\, b | \Lambda_b(p-q) \rangle
= (m_{\Lambda_b}+m_\Lambda)\, G(q^2)\,
\bar{u}_\Lambda(p)\, i\gamma_5\, u_{\Lambda_b}(p-q)\,,
\label{eq:PSff}
\end{equation}
where $G(q^2)$ is a dimensionless form factor.  

For the vector and axial-vector weak transition currents, the hadronic matrix elements are parametrised in terms of three independent form factors each. Explicitly, for the vector current one has:  
\begin{align}
\langle \Lambda(p) | \bar{s}\,\gamma_\mu\, b | \Lambda_b(p-q) \rangle
&=
\bar{u}_\Lambda(p)\left[
f_1(q^2)\,\gamma_\mu
+ i\,\frac{f_2(q^2)}{m_{\Lambda_b}}\,\sigma_{\mu\nu} q^\nu
+ \frac{f_3(q^2)}{m_{\Lambda_b}}\, q_\mu
\right] u_{\Lambda_b}(p-q)\,,
\label{eq:vecff}
\end{align}
while for the axial-vector current the corresponding decomposition reads:
\begin{align}
\langle \Lambda(p) | \bar{s}\,\gamma_\mu\gamma_5\, b | \Lambda_b(p-q) \rangle
&=
\bar{u}_\Lambda(p)\left[
g_1(q^2)\,\gamma_\mu
+ i\,\frac{g_2(q^2)}{m_{\Lambda_b}}\,\sigma_{\mu\nu} q^\nu
+ \frac{g_3(q^2)}{m_{\Lambda_b}}\, q_\mu
\right]\gamma_5\, u_{\Lambda_b}(p-q)\,.
\label{eq:axff}
\end{align}

The form factors $f_i(q^2)$ and $g_i(q^2)$ encode the local nonperturbative QCD matrix elements of the $\Lambda_b \to \Lambda$ transition and constitute important hadronic inputs for rare and semileptonic $\Lambda_b$ decays. 
Although the $\Lambda_b$ ground-state provides the leading contribution to the hadronic dispersion relation, the relatively small mass gap between the $\Lambda_b$ and its lowest negative-parity excitation $\Lambda_b^*$ implies that the latter can give a non-negligible contribution in the sum-rule analysis. It is therefore natural to include the $\Lambda_b^*$ pole explicitly in the hadronic representation of the correlation function, in order to achieve a more complete and controlled description of the intermediate-state contributions.
The negative-parity pole contribution is specified by the coupling of the interpolating current to the $\Lambda_b^*$ state and by the corresponding weak transition matrix element. We define 
\begin{eqnarray}
\bra{0}\eta^{(l)}_{\Lambda_b}\ket{\Lambda_b^*(p-q)} =
m_{\Lambda_b^*}\lambda_{\Lambda_b^*}^{(l)}\,\gamma_5 u_{\Lambda_b^*}(p-q)\,,
\label{eq:dclams}
\end{eqnarray}
where $\lambda_{\Lambda_b^*}^{(l)}$ denotes the corresponding coupling, while the explicit $\gamma_5$ accounts the opposite parity of the $\Lambda_b^*$ state. 

As an explicit example, for the pseudoscalar transition current one obtains:
\begin{equation}
\langle \Lambda(p) | m_b \,\bar{s}\, i\gamma_5\, b | \Lambda_b^*(p-q) \rangle
=
(m_{\Lambda_b^*}-m_\Lambda)\,\hat{G}(q^2)\,
\bar{u}_\Lambda(p)\, i\, u_{\Lambda_b^*}(p-q)\,,
\label{eq:PSffst}
\end{equation}
where $\hat{G} (q^2)$ denotes the pseudoscalar form factor associated with the $\Lambda_b^*$ intermediate state. Analogously, the matrix elements of the vector and axial-vector transition currents between the $\Lambda_b^*$ and $\Lambda$ baryons are parameterised in terms of form factors $\hat{f}_{1,2,3}(q^2)$ and $\hat{g}_{1,2,3}(q^2)$, respectively, whose explicit Lorentz decompositions follow the same pattern as in Eqs.~(\ref{eq:vecff}) and~(\ref{eq:axff}), up to the appropriate parity structure.

To complete the construction of the hadronic representation, we decompose the correlation function in Eq.~(\ref{eq:corrl}) into Lorentz structures, whose coefficient functions depend on the kinematic variables $(p-q)^2$ and $q^2$. These invariant amplitudes form the basis for establishing dispersion relations in the $\Lambda_b$ channel. For the pseudoscalar transition current, the correlation function can be written as:
\begin{eqnarray}
    F^{(l)}_5({p},q) = \left[F_1^{(l)}((p-q)^2,q^2)+
\slashed{q}\,F_2^{(l)} ((p-q)^2,q^2)
\,\right]i\gamma_5 u_\Lambda(p)\,,
\label{eq:pscorr}
\end{eqnarray}
where $l=\mathcal{P},\mathcal{A},\mathcal{I}$ labels the choice of the $\Lambda_b$ interpolating current.  

For the vector transition current, the correlation function is parameterised in terms of six independent invariant amplitudes as:  
\begin{eqnarray}
    F^{(l)}_\mu(p,q) =
\left(\widetilde{F}_1^{(l)}\,p_\mu\, + \widetilde{F}_2^{(l)}\, p_\mu\slashed{q}\, +
\widetilde{F}^{(l)}_3\,\gamma_\mu
\, + \widetilde{F}_4^{(l)}\,\gamma_\mu\slashed{q}\, +
\widetilde{F}_5^{(l)}\,q_\mu\, + \widetilde{F}_6^{(l)}\,q_\mu\slashed{q}\,\right)u_\Lambda(p)\,,
 \label{eq:veccorr}
\end{eqnarray}
while for the axial-vector transition current one finds:  
\begin{eqnarray}
    F^{(l)}_{\mu 5}({p},q) =
\left(\bar{F}_1^{(l)}\,{p_\mu}\,+\bar{F}_2^{(l)}\,{p_\mu}\slashed{q}\,+
\bar{F}^{(l)}_3\,\gamma_\mu
\,+\bar{F}_4^{(l)}\,\gamma_\mu\slashed{q}\,+
\bar{F}_5^{(l)}\,q_\mu\,+\bar{F}_6^{(l)}\,q_\mu\slashed{q}\,\right)
\gamma_5 u_\Lambda(p)\,.
 \label{eq:axvecorr}
\end{eqnarray}
Employing the hadronic matrix elements introduced above and summing over the spins of the intermediate $\Lambda_b$ and $\Lambda_b^*$ states, one obtains a dispersion relation for the invariant amplitudes in the variable $(p-q)^2$. As an explicit example, for the pseudoscalar transition current, the combined invariant amplitudes $F^{(l)}_{1,2}$ read:  

\begin{align}
F^{(l)}_5\big({p}, q\big) & = \frac{
m_{\Lambda_b}
{\lambda_{\Lambda_b}^{(l)}}\, G(q^2)
}{m_{\Lambda_b}^2-(p-q)^2
} \left[ (m_{\Lambda_b}^2-m_\Lambda^2) -(m_{\Lambda_b}+m_\Lambda) \slashed{q} \right]i\gamma_5 u_\Lambda(p) \nonumber \\ & + \frac{
m_{\Lambda_b^*}
{\lambda_{\Lambda_b^*}^{(l)}}\, \hat{G}(q^2)}{m_{\Lambda_b^*}^2-(p-q)^2
} \left[(m_{\Lambda_b^*}^2-m_\Lambda^2) + (m_{\Lambda_b^*}-m_\Lambda) \slashed{q} \right]i\gamma_5 u_\Lambda(p)
\nonumber\\ & +
\int_{s_0^h}^{\infty}
\frac{ds}{s-(p-q)^2} (\rho_1^{(l)}(s,q^2) + \rho_2^{(l)}(s,q^2) \slashed{q} ) i\gamma_5 u_\Lambda(p)\,,
\label{eq:psdisp}
\end{align}
where $\rho_{1,2}^{(l)}(s,q^2)$ denote the hadronic spectral densities associated with higher resonances and continuum states carrying the quantum numbers of the $\Lambda_b$ baryon, and $s_0^h$ is the corresponding effective continuum threshold.
In the case of vector and axial-vector transition currents, an analogous dispersion representation arises for each of the six invariant amplitudes defined in Eqs.~(\ref{eq:veccorr}) and (\ref{eq:axvecorr}). For brevity, the explicit expressions for these amplitudes are collected in Appendix~\ref{app:A}.

\section{Light-cone OPE and dispersion relation}
\label{sec:lcope}
We compute the correlation functions defined in Eq.~(\ref{eq:corrl}) in QCD by performing a light-cone expansion of the time-ordered product of currents at $x^2 \to 0$. The external kinematics are chosen such that the momentum flowing through the $\Lambda_b$ channel, $(p-q)^2$, is spacelike and large, ensuring that the correlation function is dominated by light-like distances and can be expressed in terms of $\Lambda$-baryon DAs.
Throughout this analysis, we neglect light-quark masses, retaining only the heavy-quark mass $m_b$ and the physical $\Lambda$-baryon mass $m_\Lambda$, which enters the kinematics and the normalisation of the $\Lambda$ DAs. At leading order in $\alpha_s$, the correlation function is expressed as a convolution of perturbatively calculable hard-scattering kernels with three-quark $\Lambda$ DAs. The light-cone expansion is organised in contributions of increasing twist. In the present work, we include all three-quark $\Lambda$ DAs from twist-3 up to twist-6. The explicit definitions of the $\Lambda$ DAs, together with their normalisation and parameters, are collected in Appendix~\ref{app:LCDAs}.

Substituting the light-cone expansion of the quark-level matrix elements in terms of $\Lambda$ DAs into the correlation functions $F^{(l)}_{5}$, $F^{(l)}_{\mu}$, and $F^{(l)}_{\mu5}$, one can project out the corresponding invariant amplitudes defined in Eqs.~(\ref{eq:pscorr})-(\ref{eq:axvecorr}). On the QCD side, these amplitudes are obtained as convolution integrals of the hard-scattering kernels with the $\Lambda$ DAs and depend on the variables $(p-q)^2$ and $q^2$.
Each invariant amplitude admits an independent dispersion representation in the $\Lambda_b$ channel. For a given choice of interpolating current, the dispersion relations contain pole contributions from the $\Lambda_b$ and $\Lambda_b^*$ states, as well as continuum contributions from higher-mass states. In the case of the vector and axial-vector transition currents, six linearly independent invariant amplitudes arise, leading to six corresponding dispersion relations.
The dispersion relations obtained for the individual invariant amplitudes generally receive contributions from both the $\Lambda_b$ ground state and its negative-parity excitation $\Lambda_b^*$. To eliminate the explicit negative-parity contribution, we form suitable linear combinations of invariant amplitudes for which the $\Lambda_b^*$ pole cancels identically. This cancellation exploits the different parity structure of the $\Lambda_b$ and $\Lambda_b^*$ pole terms, leaving the $\Lambda_b$ ground-state pole together with the higher-state and continuum contributions. 

As an explicit example, for the pseudoscalar transition current, an appropriate linear combination of the invariant amplitudes $F_1^{(l)}$ and $F_2^{(l)}$ eliminates the $\Lambda_b^*$ contribution and leads to the following dispersion relation:
 \begin{align}
 \frac{ m_{\Lambda_b} (m_{\Lambda_b} + m_\Lambda)  (m_{\Lambda_b}+
m_{\Lambda_b^*}) \lambda_{\Lambda_b}^{(l)} G(q^2)}{m_{\Lambda_b}^2-(p-q)^2}+
\int\limits_{s_0^h}^{\infty} ds \, \frac{ \rho_1^{(l)}(s,
q^2)-(m_{\Lambda_b^*} + m_\Lambda)\rho_2^{(l)}(s, q^2)}{s -(p-q)^2 }
&& \nonumber \\
 = \bigg [ F_1^{(l)}((p-q)^2, q^2)- (m_{\Lambda_b^*} + m_\Lambda
)F_2^{(l)}((p-q)^2, q^2) \, \bigg ] \,. && \label{eq:pseudosca}     
 \end{align}
The contributions of higher resonances and continuum states in the $\Lambda_b$ channel are approximated using semi-global quark-hadron duality. Concretely, the higher-state and the continuum contribution in the hadronic dispersion relation is approximated by the corresponding QCD spectral density above an effective threshold $s_0$, 
\begin{align}
\int_{s_0^h}^{\infty} ds \,
\frac{\rho_1^{(l)}(s,q^2)-(m_{\Lambda_b^*}+m_\Lambda)\rho_2^{(l)}(s,q^2)}
{s-(p-q)^2}
\approx
\frac{1}{\pi}\int_{s_0}^{\infty} ds\,
\frac{{\rm Im}_s F_1^{(l)}(s,q^2)
-(m_{\Lambda_b^*}+m_\Lambda)\,{\rm Im}_s F_2^{(l)}(s,q^2)}
{s-(p-q)^2}\,,
\label{eq:qhd}
\end{align}
where $s_0$ denotes the effective continuum threshold and ${\rm Im}_s F_{1,2}^{(l)}$ are obtained from the light-cone OPE.

Employing the decomposition of the $\Lambda$-baryon matrix elements in terms of DAs, the invariant amplitudes can be written as convolution integrals over the momentum fractions $x_i$. After using the $\delta$-function constraint to eliminate one variable and performing one trivial integration, the correlation functions reduce to the form,
\begin{equation}
F_j^{(l)} =
\frac{m_b}{4}
\sum_{n=1}^{3}
\int_0^1 dx_2\,
\frac{T^{(l)}_{jn}(x_2,(p-q)^2,q^2)}{\Delta^n}\,,
\label{eq:qcdcor}
\end{equation}
with
$\Delta = m_b^2 - x_2 (p-q)^2 + x_2\bar{x}_2 m_\Lambda^2 - \bar{x}_2 q^2$.
The integrals can be cast into a dispersion form by introducing the variable:
\begin{equation}
s(x_2)=\frac{m_b^2-\bar{x}_2 q^2 + x_2\bar{x}_2 m_\Lambda^2}{x_2}\,,
\end{equation}
with the inverse transformation:
\begin{equation}
x_2(s)=\frac{1}{2m_\Lambda^2}
\left[
m_\Lambda^2+q^2-s
+\sqrt{(s-q^2-m_\Lambda^2)^2+4m_\Lambda^2(m_b^2-q^2)}
\right].
\end{equation}
The above representation provides the QCD input needed for the construction of light-cone sum rules for the $\Lambda_b \to \Lambda$ form factors, which is discussed in the next section.

\section{Light-cone sum rules for $\Lambda_b \to \Lambda$ form factors}
\label{sec:lcsr-ff}
Combining the light-cone OPE of the correlation functions with their hadronic dispersion representations, and applying quark-hadron duality together with a Borel transformation in the $\Lambda_b$ channel, we obtain light-cone sum rules for the $\Lambda_b \to \Lambda$ transition form factors. We collect the resulting expression below.
The light-cone sum rule for the pseudoscalar form factor is:
\begin{align}
G(q^2)
&=
\frac{e^{m_{\Lambda_b}^2/M^2}}
{m_{\Lambda_b}(m_{\Lambda_b}+m_\Lambda)(m_{\Lambda_b}+m_{\Lambda_b^*})
\lambda_{\Lambda_b}^{(l)}}
\frac{1}{\pi}
\int_{m_b^2}^{s_0} ds\, e^{-s/M^2}
\nonumber\\
&\hspace{1cm}\times
\left[
{\rm Im}_s F_1^{(l)}(s,q^2)
-(m_{\Lambda_b^*}+m_\Lambda)\,
{\rm Im}_s F_2^{(l)}(s,q^2)
\right]\,.
\label{eq:psedoFF}
\end{align}

For the vector transition current, the corresponding linear combination of the invariant amplitudes gives the LCSRs for the form factors $f_{1,2,3}(q^2)$. They read:
\begin{align}
 f_1(q^2)  =\frac{e^{m_{\Lambda_b}^2/M^2}}{2 m_{\Lambda_b} (m_{\Lambda_b}+m_{\Lambda_b^*})\lambda_{\Lambda_b}^{(l)}}\frac1{\pi}\int\limits_{m_b^2}^{s_0}
ds~ e^{-s/M^2}
    \bigg [
(m_{\Lambda_b}+ m_\Lambda )\Big({\rm Im}_{s}\tilde{F}^{(l)}_1(s, q^2)
\nonumber \\
- ( m_{\Lambda_b^*} -m_\Lambda) {\rm Im}_{s}\tilde{F}^{(l)}_2(s,
q^2)\Big) + 2 {\rm Im}_{s}\tilde{F}^{(l)}_3(s, q^2)+ 2
(m_{\Lambda_b^{*}}-m_{\Lambda_b}) {\rm Im}_{s}\tilde{F}^{(l)}_4(s,
q^2)\bigg] \,, \label{eq:srf1}
\end{align}

\begin{align}
f_2(q^2) & =
\frac{e^{m_{\Lambda_b}^2/M^2}}{2(m_{\Lambda_b}+m_{\Lambda_b^*})\lambda_{\Lambda_b}^{(l)} }\frac1{\pi}\int\limits_{m_b^2}^{s_0} ds~ e^{-s/M^2}\, \Bigg [ {\rm Im}_{s}\tilde{F}^{(l)}_1(s, q^2) 
-(m_{\Lambda_b^{\ast}}- m_\Lambda )  {\rm Im}_{s}\tilde{F}^{(l)}_2(s,
q^2) \nonumber\\ &\qquad  - 2
 {\rm Im}_{s}\tilde{F}^{(l)}_4(s, q^2)  \Bigg]\, ,
\label{eq:vecf2}
\end{align}
and 
\begin{align}
f_3(q^2) & =
\frac{e^{m_{\Lambda_b}^2/M^2}}{2(m_{\Lambda_b}+m_{\Lambda_b^*})\lambda_{\Lambda_b}^{(l)} }\frac1{\pi}\int\limits_{m_b^2}^{s_0} ds~ e^{-s/M^2}\, \Bigg [ {\rm Im}_{s}\tilde{F}^{(l)}_1(s, q^2)  
  -(m_{\Lambda_b^{\ast}}- m_\Lambda ) \big(2\ {\rm Im}_{s}\tilde{F}^{(l)}_6(s,
q^2)\nonumber \\&\qquad + {\rm Im}_{s}\tilde{F}^{(l)}_2(s,
q^2) \big)  +2  \big( 
 {\rm Im}_{s}\tilde{F}^{(l)}_5(s, q^2) + \ {\rm Im}_{s}\tilde{F}^{(l)}_4(s, q^2) \big) \Bigg]\, .
\label{eq:vecf3}
\end{align}
The LCSRs for the axial-vector form factors $g_1(q^2)$, $g_2(q^2)$, and $g_3(q^2)$ follow from Eqs.~\eqref{eq:srf1}--\eqref{eq:vecf3} by replacing the vector invariant amplitudes ${\rm Im}_s \tilde{F}_j^{(l)}$ by their axial-vector counterparts ${\rm Im}_s \bar{F}_j^{(l)}$ and making the replacement $m_\Lambda \to -m_\Lambda$.

In practice, the transformation of the light-cone expressions into dispersion integrals, the subtraction of continuum contributions, and the Borel transformation in the $\Lambda_b$ channel can be implemented simultaneously through a set of standard substitution rules. For the structure appearing in Eq.~(\ref{eq:qcdcor}), these read:
\begin{eqnarray}
\int dx \frac{T(x) }{\Delta} &\to&
\int\limits_{x_0}^1 \frac{dx}{x}T(x)
\exp\left(-\frac{s(x)}{M^2}\right)\,,
\nonumber
\\
\int dx \frac{T(x) }{\Delta^2} &\to&
\frac{1}{M^2} \int\limits_{x_0}^1 \frac{dx}{x^2} T(x)
\exp\left(-\frac{s(x)}{M^2}\right)
+  \frac{T(x_0)\,e^{-s_0 /M^2} }{m_b^2+x_0^2 m_\Lambda^2-q^2} \, , \nonumber \\
\int dx \frac{T(x) }{\Delta^3} &\to& \frac{1}{2
M^4} \int\limits_{x_0}^1 \frac{dx}{x^3}T(x)\exp\left(-\frac{s(x)}{M^2}\right)
+
 \frac{1}{2 M^2}  \frac{T(x_0)\,e^{-s_0 /M^2} }{x_0\left(m_b^2+x_0^2 m_\Lambda^2-q^2\right)}
 \nonumber \\  &&-   \frac12  \frac{x_0^2\, e^{-s_0 /M^2}}{(m_b^2+x_0^2 m_\Lambda^2-q^2)}
\frac{d}{dx}\left(
\frac{T(x)}{x\left(m_b^2+x^2 m_\Lambda^2-q^2\right)}
\right)\Bigg|_{x=x_0}\;.
\label{eq:intT}
\end{eqnarray}
Here $x_0 \equiv x(s_0)$ denotes the solution of $s(x_0)=s_0$, and $T(x)$ represents a generic numerator function arising from the convolution of hard-scattering kernels with $\Lambda$ DAs. The surface terms appearing on the right-hand side originate from reducing the higher-power denominators, $\Delta^{-n}$ with $n=2,3$, to the standard dispersion representation in $s-(p-q)^2$.

\section{Numerical Analysis}
\label{sec:result}

We begin by specifying the numerical values of input parameters entering the LCSRs. The parameters of the $\Lambda$-baryon DAs are taken from QCD sum-rule analyses and are summarised in Table~\ref{tab:input parameters}. The explicit definitions and conventions for the $\Lambda$ DAs are collected in {Appendix~\ref{app:LCDAs}}. The baryon masses are taken from the Particle Data Group~\cite{ParticleDataGroup:2024cfk}.
\begin{table}[t]
    \renewcommand{\arraystretch}{1.5}
    \centering
    \begin{tabular}{|c|>{\centering\arraybackslash}p{7cm}|>{\centering\arraybackslash}p{2cm}|}
         \hline
         Parameter & Value(s) & Reference(s) \\ \hline \hline
         $G_F$ & $1.1663787(6)\times 10^{-5}\ {\rm GeV}^{-2}$ & \cite{ParticleDataGroup:2024cfk} \\
        $\alpha_{\rm em}^{-1}$ & $137.035999084(21)$ & \cite{ParticleDataGroup:2024cfk} \\
        $|V_{tb}V_{ts}^*|$ & $0.0405 \pm 0.0008$ & \cite{ParticleDataGroup:2024cfk}  \\
        $\tau_{\Lambda_b}$ & $(1.470 \pm 0.009)\times 10^{-12}\ {\rm s}$ & \cite{ParticleDataGroup:2024cfk} \\
        $m_W$ & $(80.37 \pm 0.01)\ {\rm GeV}$ & \cite{ParticleDataGroup:2024cfk} \\
        $m_t$ & $(172.6 \pm 0.3)\ {\rm GeV}$ & \cite{ParticleDataGroup:2024cfk} \\
         $f_\Lambda$ & $(6.0 \pm 0.3) \times 10^{-3}\  {\rm GeV^2}$ & \cite{Liu:2008yg} \\ 
         $\lambda_1$ & $(1.0 \pm 0.3) \times 10^{-2}\ {\rm GeV^2}$ & \cite{Liu:2008yg} \\ 
         $\lambda_2$ & $(0.83 \pm 0.05) \times 10^{-2} {\rm GeV^2}$ & \cite{Liu:2008yg} \\ 
         $\lambda_3$ & $(0.83 \pm 0.05) \times 10^{-2} {\rm GeV^2}$ & \cite{Liu:2008yg} \\ 
         $m_\Lambda$ & $1.115 \, {\rm GeV}$ &\cite{ParticleDataGroup:2024cfk}  \\ 
         $m_{\Lambda_b}$ & $5.620 \, {\rm GeV}$ & \cite{ParticleDataGroup:2024cfk} \\ 
         $m_{\Lambda_b^*} $ & 
         $ 5.912\, {\rm GeV}$  & \cite{ParticleDataGroup:2024cfk}\\
         $\lambda_{\Lambda_b}^{\mathcal{P}} $ & 
         $ (0.0109 \pm 0.0031)\, {\rm GeV}^2$  & \cite{Khodjamirian:2011jp}\\
       $\lambda_{\Lambda_b}^{\mathcal{A}} $ & $ (0.0127 \pm 0.0034)\, {\rm GeV}^2$  & \cite{Khodjamirian:2011jp}\\
       $\lambda_{\Lambda_b}^{\mathcal{I}} $ & $ (0.022 \pm 0.0049)\, {\rm GeV}^2$  & \cite{Khodjamirian:2011jp} \\
       $M^2 $ & $ (20 \pm 5)\, {\rm GeV}^2$  & [see text]\\
       $s_0$ & $ 40 \pm 1\, {\rm GeV}^2$  & [see text]\\
\hline
    \end{tabular}
    \caption{Numerical inputs used in the form-factor LCSRs and in the illustrative differential branching-fraction calculation. The Borel parameter quoted here is the form-factor LCSR parameter $M^2_{\rm LCSR}\equiv M^2$; the distinct two-point-sum-rule window used to determine $s_0$ is specified in the text.  }
    \label{tab:input parameters}
\end{table}
For the heavy-quark mass entering the correlation functions, we employ the $\overline{\rm MS}$ mass scheme, with $\bar{m}_b(\bar{m}_b)=4.16\pm0.03~{\rm GeV}$~\cite{Chetyrkin:2009fv}. The present LCSRs are evaluated at leading order in $\alpha_s$. 
To account for the uncalculated perturbative corrections, we assign an additional $50\%$ relative uncertainty to each form factor. The size of this estimate is motivated by the sizable radiative effects found in ~\cite{Wang:2015ndk} for $\Lambda_b\to\Lambda$ form factors in a different LCSR
formulation. We treat this contribution as a multiplicative uncertainty fully correlated in $q^2$ for a given form factor/current combination,
\begin{equation}
(\mathrm{Cov}_{\alpha_s})_{ij}
=
(0.50)^2 F(q_i^2)F(q_j^2).
\end{equation} 
No correlation of this missing-perturbative uncertainty is assumed between different form factors or different interpolating currents.
For each interpolating current, we use the corresponding
$\Lambda_b$ coupling $\lambda_{\Lambda_b}^{(l)}$ obtained from the
two-point QCD sum-rule analysis of Ref.~[44], with the numerical values given in Table~\ref{tab:input parameters}.

\begin{figure}[t]
    \centering
    \begin{subfigure}{.45\textwidth}
        \includegraphics[height=0.2\textheight, keepaspectratio]{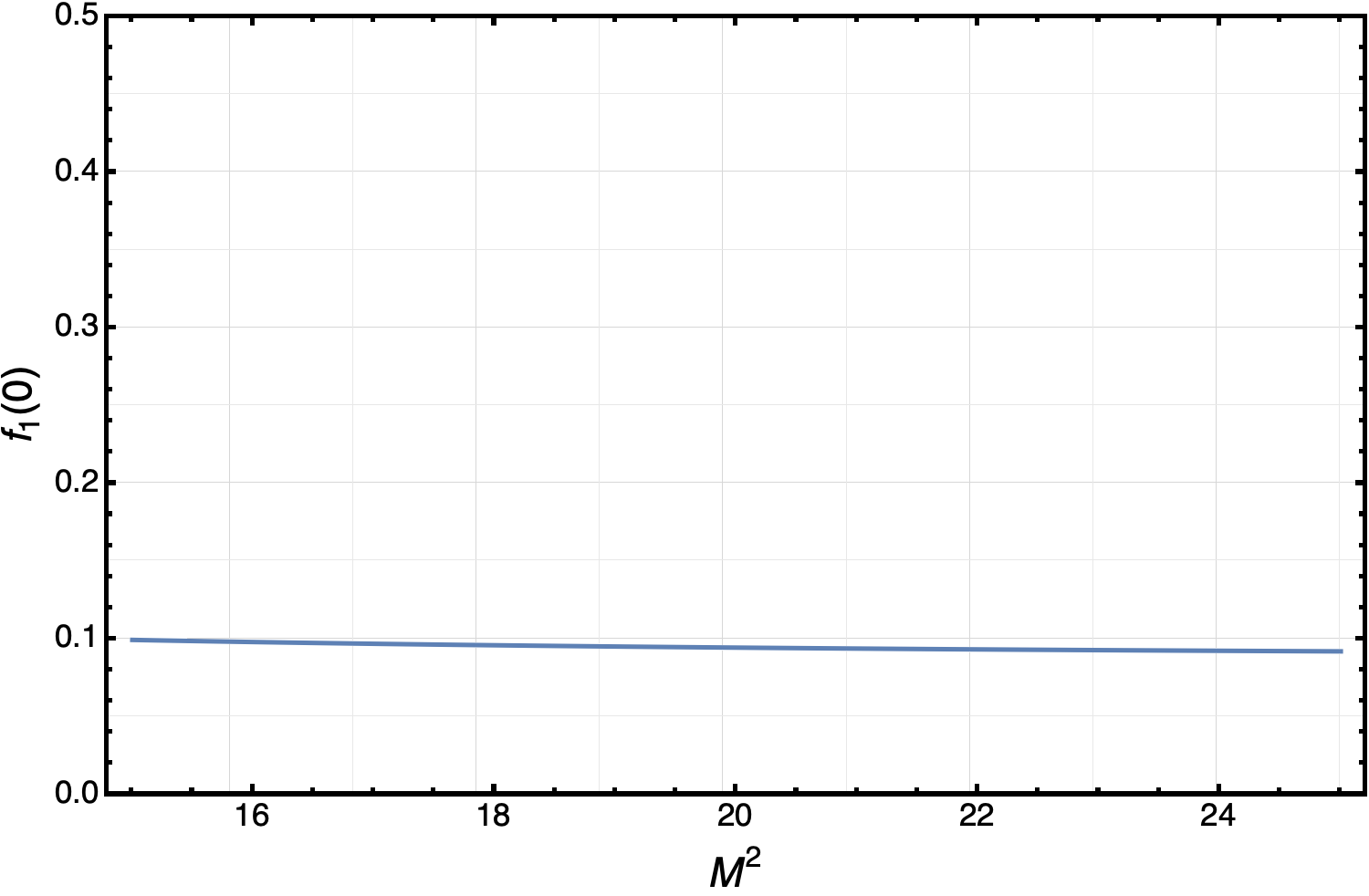}
    \caption{}
    \label{fig:stability}
    \end{subfigure}
    \begin{subfigure}{.45\textwidth}
        \includegraphics[height=0.2\textheight, keepaspectratio]{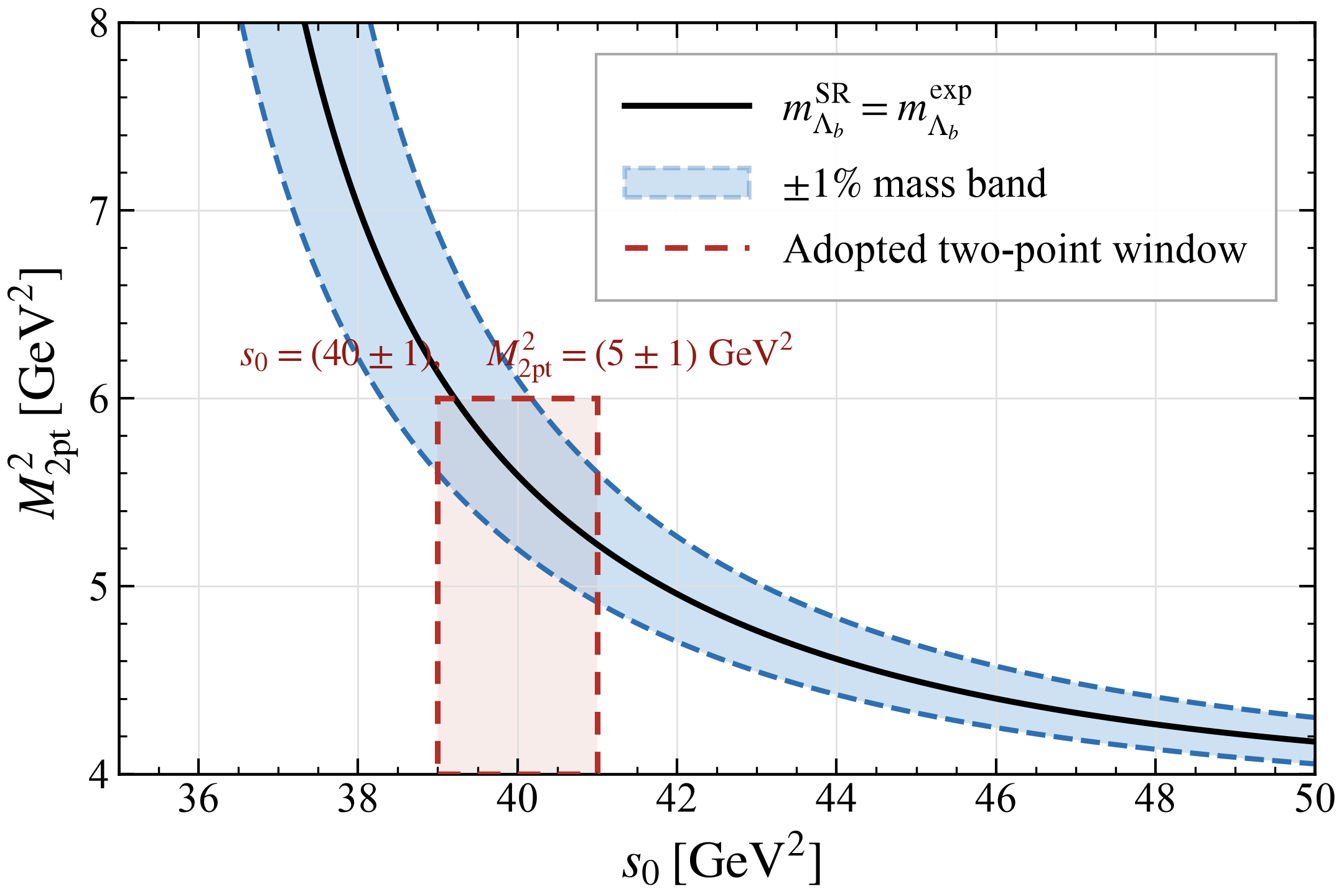}
    \caption{}
    \label{fig:stability2}
    \end{subfigure}
    \caption{(a) Dependence of $f_1(0)$ on the form factor Borel parameter $M_{\rm LCSR}^2$ for the pseudoscalar interpolating current. (b) $\Lambda_b$ mass extracted from the two-point sum rule in the
$(s_0,M_{\rm 2pt}^2)$ plane. The central contour corresponds to the
experimental mass, the adjacent contours delimit the $\pm1\%$ region, and the rectangle indicates the adopted range. }
    \label{fig:stab}
\end{figure}

The continuum threshold is determined from the corresponding two-point $\Lambda_b$ sum rule. We use the Borel window $ M_{\rm 2pt}^2=(5\pm1)~{\rm GeV}^2$ and find that $s_0=(40\pm1)~{\rm GeV}^2$ reproduces the physical $\Lambda_b$ mass within approximately $1\%$ over the adopted region, as shown in Fig.~\ref{fig:stability2}. The form factor LCSRs involve a different correlation function and are evaluated in the separate Borel window $M_{\rm LCSR}^2=(20\pm5)~{\rm GeV}^2$. The two Borel parameters therefore need not coincide. Within this window, the form factors exhibit only a mild dependence on $M_{\rm LCSR}^2$; Fig.~\ref{fig:stability} shows $f_1(0)$ for the pseudoscalar interpolating current as a representative example. We find analogous Borel stability for the remaining form factors and for all three interpolating currents. 

\begin{table}[t]
    \renewcommand{\arraystretch}{1.5}
    \centering
    \begin{tabular}{|c|c|c|c|c|}
         \hline
         Form factor & $ \eta^{(\mathcal{P})}$ & $\eta^{(\mathcal{A})}$ & $ \eta^{(\mathcal{I})}$    \\ \hline \hline
         $G(0)$ & 
         {$0.085^{+0.053}_{-0.047}$} & {$0.023^{+0.018}_{-0.016}$} & {$-0.099^{+0.052}_{-0.055}$}  \\ \hline
         $f_1(0)$ & 
        {$0.096^{+0.060}_{-0.053}$} & {$0.046^{+0.029}_{-0.025}$} & {$0.105^{+0.061}_{-0.056}$}   \\ \hline
         $f_2(0)$ & 
         {$-0.034^{+0.019}_{-0.021}$} & {$0.023^{+0.015}_{-0.013}$} & {$0.083^{+0.048}_{-0.044}$} \\ \hline
         $g_1(0)$ & 
        {$0.095^{+0.059}_{-0.052}$} & {$0.053^{+0.034}_{-0.029}$} & {$-0.058^{+0.031}_{-0.035}$}  \\ \hline
         $g_2(0)$ & 
         {$-0.014^{+0.008}_{-0.009}$} &  {$-0.062^{+0.034}_{-0.039}$} & {$-0.006^{+0.007}_{-0.008}$}  \\ \hline
    \end{tabular}
    \caption{
$\Lambda_b\to\Lambda$ form factors at $q^2=0$ obtained with the
pseudoscalar, axial-vector, and Ioffe interpolating currents. The central values correspond to the medians of the sampled distributions, and the quoted uncertainties denote the intervals between the $16^{\rm th}$ and $84^{\rm th}$ percentiles. They include the LCSR input uncertainties together with the additional correlated $50\%$ estimate of the missing perturbative corrections.}
    \label{tab:current_dependence}
\end{table}

To assess the dependence on the choice of $\Lambda_b$ interpolating current, we evaluate the form factors using the pseudoscalar, axial-vector, and Ioffe currents. For each case, the corresponding coupling $\lambda_{\Lambda_b}^{(l)}$ is sampled from the input given in Table~\ref{tab:input parameters}. All three choices lead to numerically stable LCSRs within the same Borel and continuum-threshold windows.
The central values nevertheless show a noticeable dependence on the
interpolating current. Among the three choices, the pseudoscalar current gives the closest agreement with the expected EOM and HQS relations, while larger deviations are found for the axial-vector and Ioffe currents. Once the additional uncertainty associated with the missing perturbative corrections is included, the enlarged uncertainties substantially weaken the discriminating power of the EOM/HQS relations, and no statistically significant current selection can be made at the present accuracy.
For the subsequent presentation of the $q^2$ dependence and the
phenomenological application, we use the pseudoscalar-current results. This choice is motivated by their closer agreement with the symmetry and EOM expectations at the level of central values and should not be interpreted as an exclusion of the axial-vector or Ioffe currents.

Since the perturbative $O(\alpha_s)$ corrections to the present LCSRs have not been calculated, the current dependence observed at leading order should not be overinterpreted. The additional uncertainty introduced above provides a conservative estimate of this missing perturbative effect. A calculation of the radiative corrections within the present LCSR framework would be required for a more quantitative assessment of the residual interpolating current dependence.

As a cross-check of the implementation, we replace the $\Lambda$ baryon and its DAs by the corresponding proton quantities. In this setup, we reproduce the nucleon LCSR results of~\cite{Khodjamirian:2011jp}, providing an independent check of the analytical expressions and their numerical implementation.

To examine the numerical hierarchy of the light-cone expansion, we separate the twist-3, twist-$4$, twist-$5$, and twist-$6$ contributions at $q^2=0$ for the pseudoscalar interpolating current. The results are given in Table~$3$. The hierarchy is not monotonic in twist: twist~$4$ dominates $G$, $f_1$, $f_2$, and $g_1$, whereas the twist-$5$ contribution is particularly important for $g_2$ and is also non-negligible for $f_2$. The twist-3 terms
are numerically small in this decomposition. In all cases the twist-6 contribution is strongly suppressed.
We have also checked the corresponding decomposition for the axial-vector and Ioffe interpolating currents. The relative sizes of the lower-twist terms depend on the choice of current and form factor, while the twist-$6$ contribution remains numerically small.

\begin{table}[t]
    \renewcommand{\arraystretch}{1.5}
    \centering
    \begin{tabular}{|c|c|c|c|c|c|}
         \hline
    Form factor  & \text{twist-3} & \text{twist-4} & \text{twist-5} & \text{twist-6} & \text{total} \\ \hline
\hline
$G(0)$   & $0.00021$ & $0.09292$ & $-0.00964$ & $0.00031$ & $0.08381$\\
$f_1(0)$ & $0.00020$ & $0.09573$ & $-0.00179$ & $0.00030$ & $0.09444$\\
$f_2(0)$ & $-0.00012$ & $-0.04356$ & $0.01049$ & $-0.00018$ & $-0.03336$\\
$g_1(0)$ & $0.00020$ & $0.09333$ & $-0.00060$ & $0.00030$ & $0.09323$\\
$g_2(0)$ & $-0.00012$ & $-0.00495$ & $-0.00868$ & $-0.00018$ & $-0.01392$\\
\hline
    \end{tabular}
    \caption{Individual twist-3, twist-4, twist-5, and twist-6 contributions to the $\Lambda_b\to\Lambda$ form factors at $q^2=0$ for the pseudoscalar interpolating current, evaluated at the central input values with $m_{\Lambda_b^*}=5.912~{\rm GeV}$. The entries are signed contributions, and their sum gives the total in the last column.}
    \label{tab:tw_contri}
\end{table}
The entries in Table~\ref{tab:tw_contri} are evaluated at fixed central inputs and therefore should not be identified with the medians quoted in Table~2, which are obtained from the full parameter scan. We do not assign separate uncertainties to the individual twist contributions, since they depend on common input parameters and are correlated. Their signed sum reproduces the fixed-input total shown in the last column.

\begin{figure}[h!]
    \centering
    \includegraphics[width=0.8\linewidth]{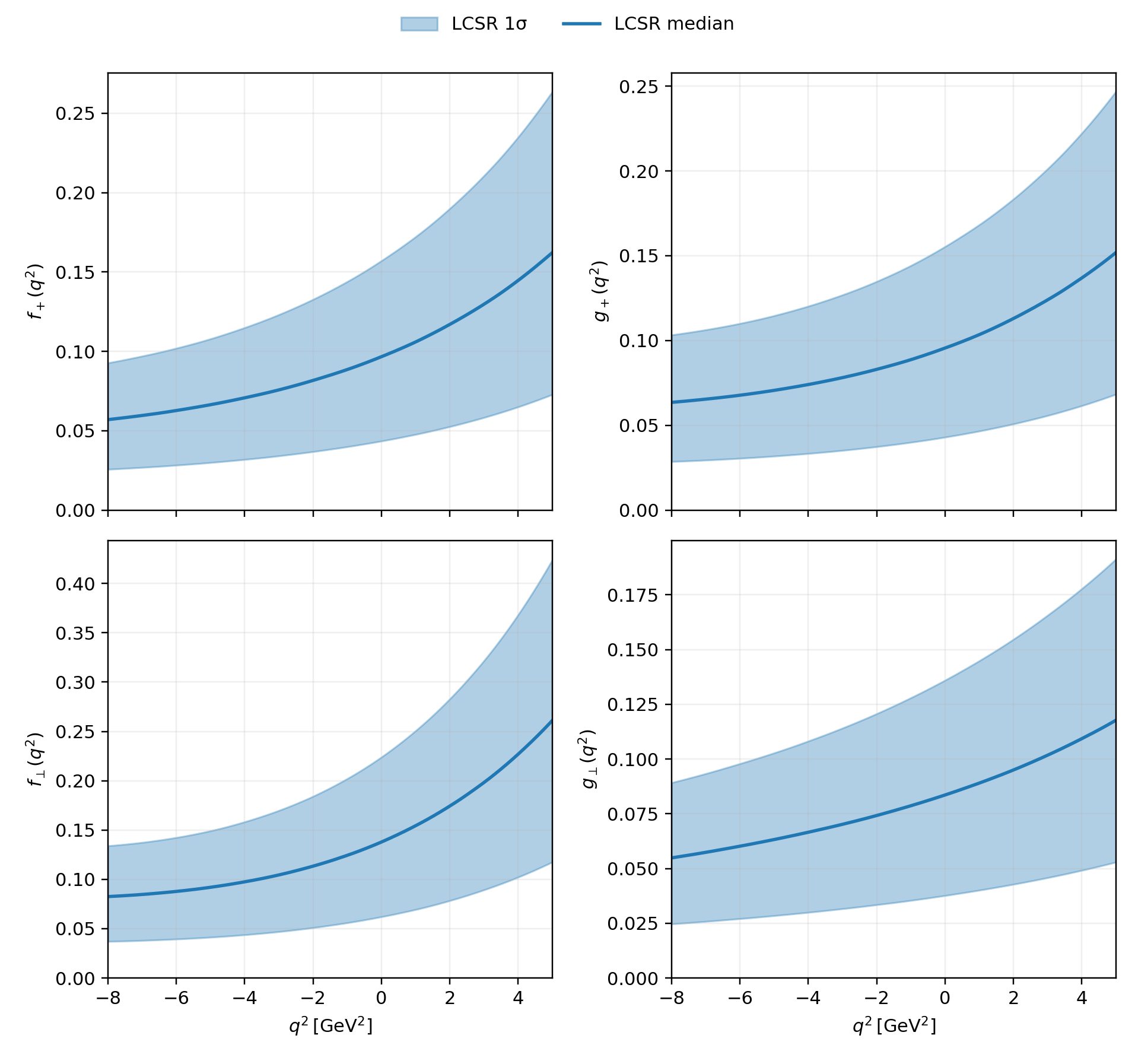}
    \caption{Helicity form factors $f_+$, $f_\perp$, $g_+$, and $g_\perp$ for the $\Lambda_b\to\Lambda$ transition obtained from the pseudoscalar-current LCSRs. The solid lines denote the median values, while the shaded bands represent the total LCSR uncertainties, including the correlated estimate of missing $O(\alpha_s)$ corrections.}
\label{fig:ff_lcsr}
\end{figure}

The LCSRs derived in Sec.~\ref{sec:lcsr-ff} determine the form factors $f_{1,2,3}(q^2)$ and $g_{1,2,3}(q^2)$ in the Dirac basis. For phenomenological applications and comparison with lattice-QCD results, we transform them to helicity basis. For massless leptons, only
$f_+$, $f_\perp$, $g_+$, and $g_\perp$ contribute to the vector and
axial-vector helicity amplitudes~\cite{Detmold:2016pkz,Feldmann:2011xf}. Their relation to the Dirac-basis form factors is 
\begin{align}
f_+(q^2) &= f_1(q^2) - \frac{q^2}{m_{\Lambda_b}(m_{\Lambda_b}+m_\Lambda)}\, f_2(q^2)\,, \\
f_\perp(q^2) &= f_1(q^2) - \frac{(m_{\Lambda_b}+m_\Lambda)}{m_{\Lambda_b}}\, f_2(q^2)\,, \\
g_+(q^2) &= g_1(q^2) + \frac{q^2}{m_{\Lambda_b}(m_{\Lambda_b}-m_\Lambda)}\, g_2(q^2)\,, \\
g_\perp(q^2) &= g_1(q^2) + \frac{(m_{\Lambda_b}-m_\Lambda)}{m_{\Lambda_b}}\, g_2(q^2)\,,
\end{align}
The remaining timelike helicity form factors, which involve
$f_3(q^2)$ and $g_3(q^2)$, do not contribute in the massless-lepton limit considered here.

Fig.~\ref{fig:ff_lcsr} shows the helicity form factors obtained from the pseudoscalar-current LCSRs. The solid curves denote the median values and the bands the corresponding central $68\%$ intervals, including the additional uncertainty associated with missing perturbative corrections.
To parametrise their $q^2$ dependence, we employ the truncated
$z$-expansion~\cite{Bourrely:2008za} 
\begin{equation}
F(q^2)=\frac{1}{1-q^2/M_F^2}\,
\sum_{n=0}^{K} a_n^{(F)}\, z(q^2)^n\,,
\qquad
z(q^2)=\frac{\sqrt{t_+-q^2}-\sqrt{t_+-t_0}}
{\sqrt{t_+-q^2}+\sqrt{t_+-t_0}}\,,
\label{eq:zparam}
\end{equation}
where $t_0=(m_{\Lambda_b}-m_\Lambda)^2$ and $t_+=(m_B+m_K)^2$. The same definition of $z(q^2)$ is used both in determining the fit coefficients and in all subsequent evaluations of the fitted form factors.
The pole masses are chosen as $M_{f_+}=M_{f_\perp}=5.416~\mathrm{GeV}$ and
$M_{g_+}=M_{g_\perp}=5.750~\mathrm{GeV}$, following the conventions adopted in lattice-QCD analyses~\cite{Detmold:2016pkz}. We retain terms through $z^2$ ($K=2$); the resulting coefficients are
listed in Table~\ref{tab:zfit}. 
\begin{table}[t]
\centering
\renewcommand{\arraystretch}{1.3}
    \small
    \begin{tabular}{|c|c|c|c|}
         \hline
         Form factor & $a_0$ & $a_1$ & $ a_2$    \\ \hline \hline
         $f_+$ & 
         {$0.625 \pm 0.405$} &
         {$-3.671 \pm 2.378$} &
         {$6.159 \pm 3.986$} \\ \hline
         $f_\perp$ & 
         {$1.409 \pm 0.910$} &
         {$-9.348 \pm 6.040$} &
         {$16.889 \pm 10.920$} \\ \hline
         $g_+$ & 
         {$0.439 \pm 0.284$} &
         {$-2.352 \pm 1.523$} &
         {$3.904 \pm 2.526$} \\ \hline
         $g_\perp$ & 
         {$0.439 \pm 0.284$} &
         {$-2.453 \pm 1.588$}&
         {$4.130 \pm 2.673$}\ \\ \hline
    \end{tabular}
\caption{Coefficients of the $K=2$ $z$-expansion for the helicity form factors obtained from the pseudoscalar-current LCSRs. The quoted uncertainties include the LCSR input uncertainties and the correlated $50\%$ estimate of missing perturbative corrections. The corresponding coefficient correlation matrices are provided in the ancillary material. }
\label{tab:zfit}
\end{table}

The LCSR uncertainties are obtained by Monte Carlo sampling of the input parameters listed in Table~\ref{tab:input parameters}. For each sampled parameter set, the complete sum rule is evaluated at the chosen $q^2$ points. We quote the median as the
central value and the central $68\%$ interval as the uncertainty. Correlations among the different $q^2$ points are retained in the subsequent $z$-expansion fits. For each form-factor/current block, the estimate of the missing perturbative correction is added as the correlated rank-one covariance contribution defined above.

For comparison with previous LCSR determinations, Table~5 collects the helicity form factors $f_+(0)$ and $g_+(0)$ from two representative calculations. At $q^2=0$, these coincide with $f_1(0)$ and $g_1(0)$ in our conventions, which avoids ambiguities associated with the definitions and sign conventions of the subleading form factors.

\begin{table}[h]
\centering
\renewcommand{\arraystretch}{1.5}
\begin{tabular}{lccc}
\hline\hline
 LCSR setup & $f_+(0)$ & $g_+(0)$ & Ref. \\
\hline
 $\Lambda$ DAs, twist $3$--$6$
&  {$0.096^{+0.060}_{-0.053}$}
&  {$0.095^{+0.059}_{-0.052}$} & This work
 \\

$\Lambda$ DAs, twist $3$--$6$
& $0.150^{+0.020}_{-0.020}$
& $0.150^{+0.020}_{-0.020}$ & \cite{Wang:2008sm}
\\

$\Lambda_b$ DAs, NLL
& $0.18^{+0.05}_{-0.06}$
& $0.18^{+0.05}_{-0.06}$ & \cite{Wang:2015ndk}
 \\
\hline\hline
\end{tabular}
\caption{Comparison of $f_+(0)$ and $g_+(0)$ with representative LCSR determinations. \cite{Wang:2008sm} employs light-$\Lambda$ DAs through twist~6, whereas \cite{Wang:2015ndk} uses $\Lambda_b$ DAs and includes perturbative/NLL corrections. For~\cite{Wang:2015ndk}, the uncertainty shown here is obtained by combining the separately quoted theory uncertainties in quadrature.
The uncertainties of the present work include the additional estimate of missing $O(\alpha_s)$ corrections.
}
\label{tab:LCSRcomparison}
\end{table}
The present results are somewhat lower in central value than the two previous determinations, but are compatible with them within the quoted uncertainties. The comparison also illustrates the residual dependence of baryonic LCSR predictions on the choice of correlation function, distribution amplitudes, and perturbative treatment.
\subsection{Combined LCSR and lattice-QCD form factors for $\Lambda_b \to \Lambda$}
\label{subsec:lcsrlatticeff}
\begin{figure}[t!]
    \centering
    \includegraphics[width=0.8\linewidth]{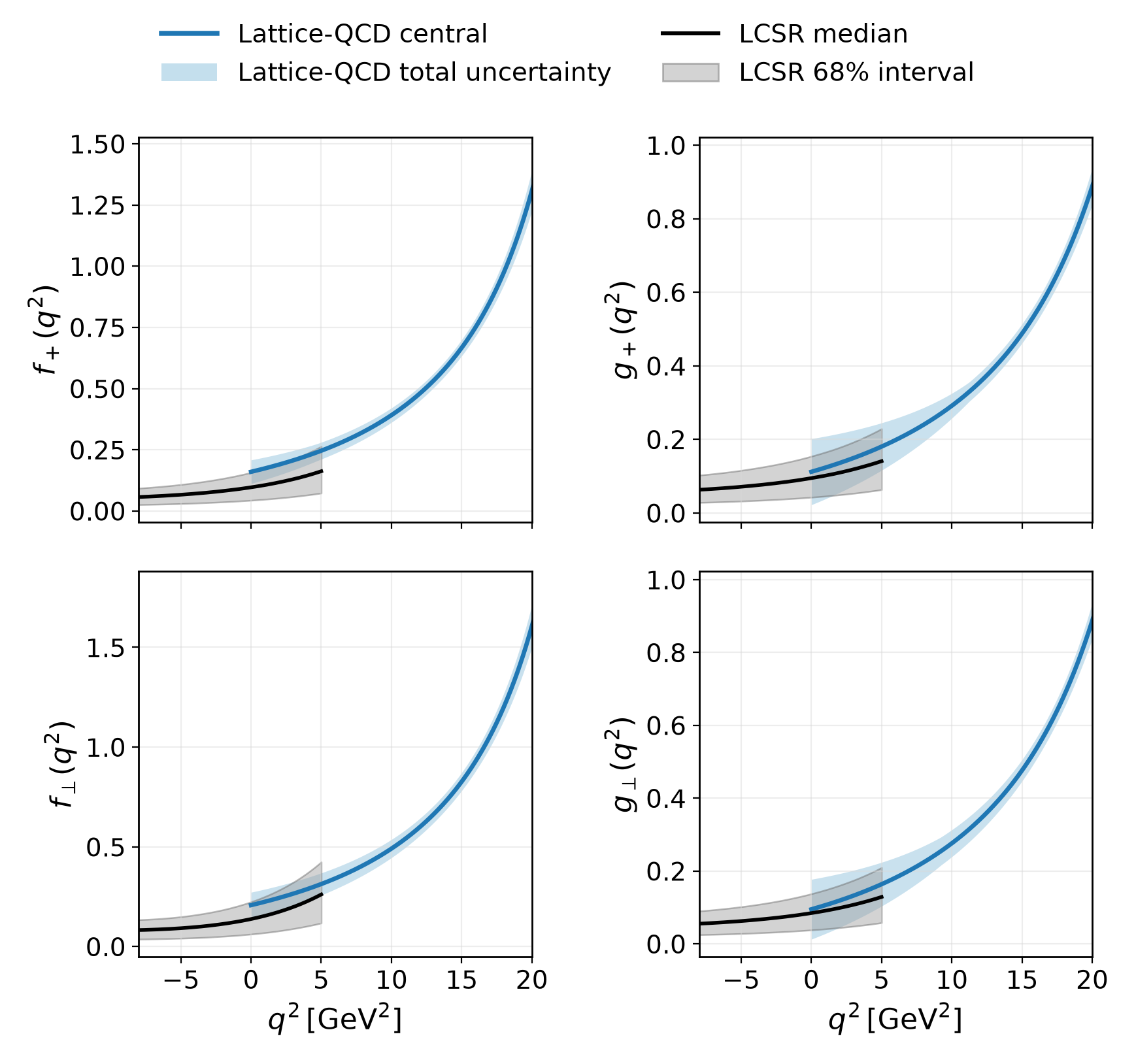}
    \caption{
    Comparison of helicity form factors $f_+(q^2)$, $g_+(q^2)$, $f_\perp(q^2)$, and
    $g_\perp(q^2)$ from LCSR (low $q^2$) and lattice-QCD (high $q^2$), shown as
    independent determinations with their respective uncertainty bands. This figure
    serves as an input-level cross-check prior to the combined fit.
    }
    \label{fig:ff_inputs}
\end{figure}
\begin{figure}[t!]
    \centering
    \includegraphics[width=0.75\linewidth]{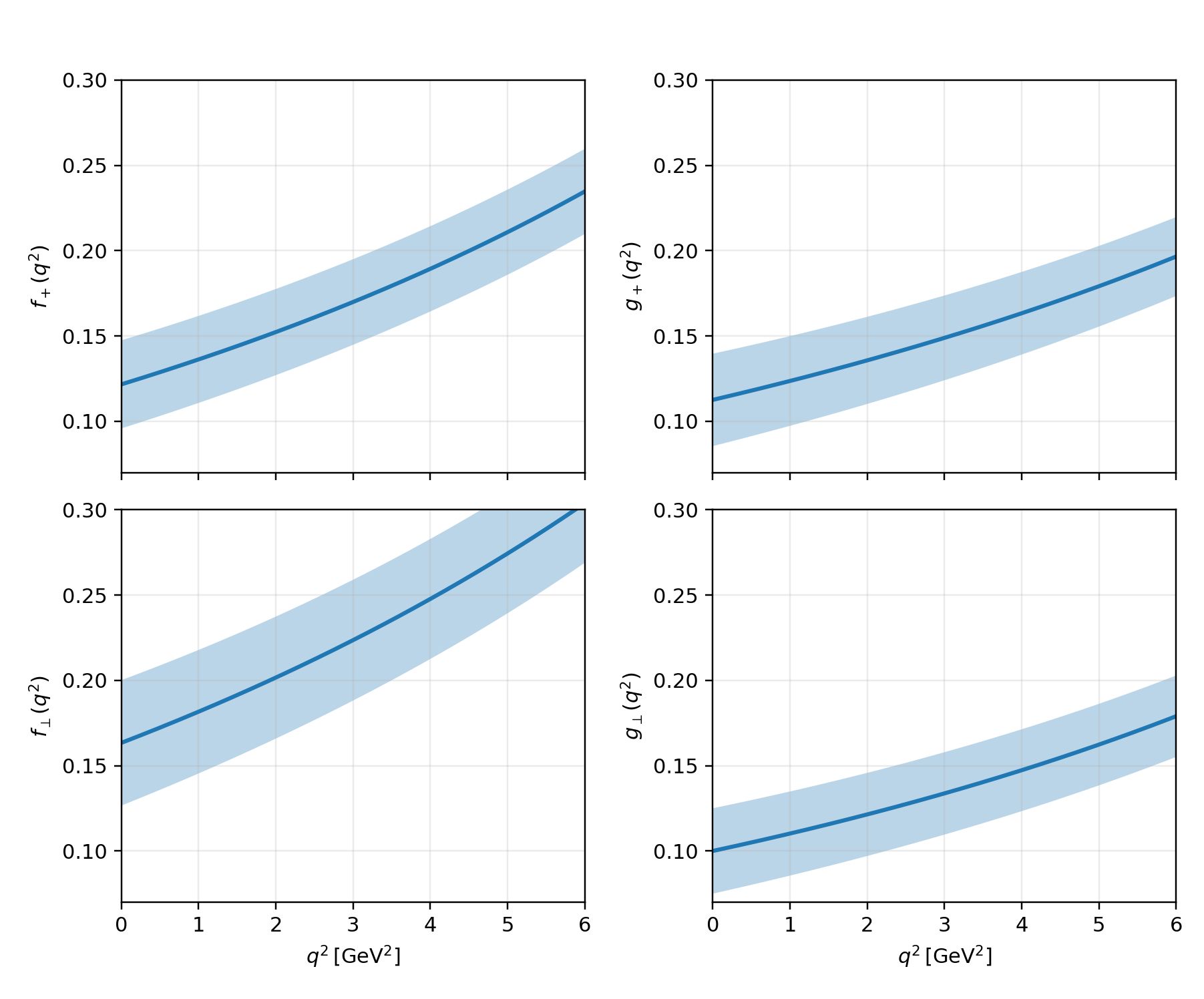}
    \caption{
   Helicity form factors $f_+$, $f_\perp$, $g_+$, and $g_\perp$ in the low-$q^2$ region obtained from the combined LCSR+lattice-QCD fit. The solid curves denote the fitted central values and the shaded bands the corresponding fit uncertainties. The LCSR contribution to the fit covariance includes the correlated estimate of missing perturbative corrections.
    }
    \label{fig:ff_combined_lowq2}
\end{figure}

To obtain a common parametrisation of the $\Lambda_b \to \Lambda$ form factors over the physical kinematic range, we perform a combined fit to the LCSR results at large recoil and the lattice-QCD results near zero recoil. The two calculations provide complementary information in well-separated kinematic regions.
Fig.~\ref{fig:ff_inputs} compares the LCSR and lattice-QCD determinations before imposing a common parametrisation. We subsequently fit both sets of inputs simultaneously using the $z$-expansion of Eq.~\eqref{eq:zparam}, shown in Fig.~\ref{fig:ff_combined_lowq2}. 

For each helicity form factor, the fit uses the four LCSR points
$q^2=\{-8,-4,-2,0\}\ {\rm GeV}^2$ and the four lattice-QCD points $q^2=\{13,15,17,19\}\ {\rm GeV}^2$. We retain the $K=2$ truncation of Eq.~\eqref{eq:zparam}. The LCSR covariance includes the additional missing-perturbative contribution described above, while the lattice-QCD covariance is taken from~\cite{Detmold:2016pkz}. The resulting
coefficients and correlation matrices are provided in the ancillary files.
The combined fit is used here to provide a common parametrisation and to compare the LCSR and lattice-QCD determinations. For the phenomenological illustration in Sec.~\ref{subsec:diffBR}, we instead retain the pseudoscalar current LCSR determination of the vector and axial-vector form factors, in order to display directly the uncertainty of the LCSR calculation presented
in this work.
\subsection{Differential branching fraction for $\Lambda_b \to \Lambda \ell^+ \ell^-$}
\label{subsec:diffBR}
We now consider the differential branching fraction for
$\Lambda_b\to\Lambda\ell^+\ell^-$ as an illustrative application of the form factors obtained above. We restrict the discussion to the low-$q^2$ region, where the LCSR calculation is directly applicable.

For the vector and axial-vector contributions we use the pseudoscalar current LCSR helicity form factors $f_+$, $f_\perp$, $g_+$, and $g_\perp$. Since tensor form factors are not calculated in the present LCSR analysis, we use the lattice-QCD results of~\cite{Detmold:2016pkz} for $h_+$, $h_\perp$, $\widetilde h_+$, and $\widetilde h_\perp$. The resulting branching-fraction calculation is therefore intended as an illustrative phenomenological application rather than a self-contained SM prediction.

\begin{figure}[t!]
    \centering
    \includegraphics[width=0.6\textwidth]{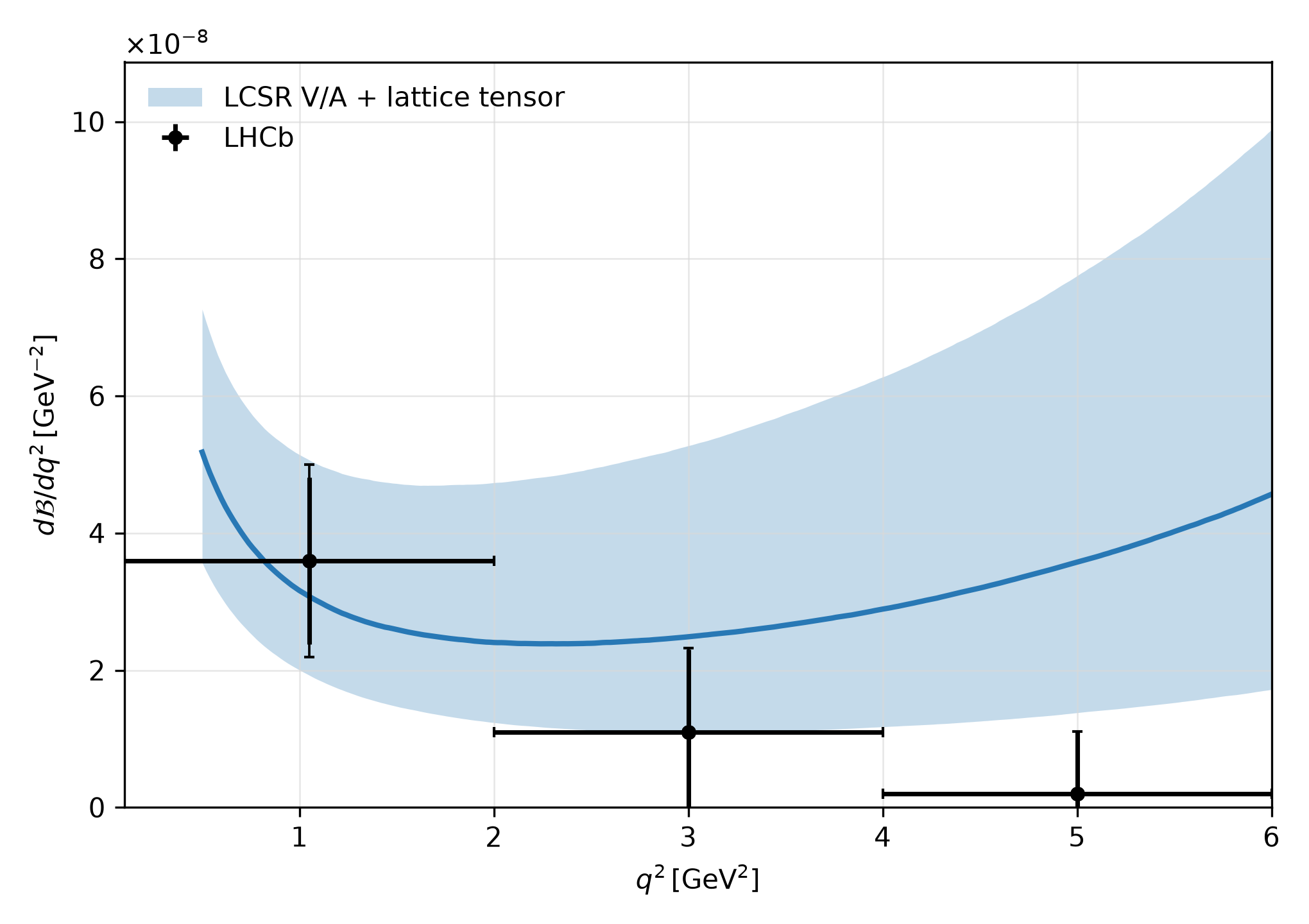}
    \caption{Illustrative differential branching fraction $d\mathcal B/dq^2$ for the decay $\Lambda_b \to \Lambda \ell^+ \ell^-$ in the low-$q^2$ region. The solid line denotes the median prediction, and the shaded band represents the propagated LCSR and lattice-QCD form factor uncertainties.}
    \label{fig:dBdq2_1_6}
\end{figure}
For the numerical evaluation we use the massless-lepton expression given in Appendix~\ref{app:diff_decay_width}. The Wilson coefficients are evaluated at $\mu_b=4.8\,{\rm GeV}$~\cite{Mahmoudi:2007vz,Mahmoudi:2008tp,Neshatpour:2021nbn}. In particular, we use
$$ C_7^{\rm eff}=-0.304,\qquad
C_9=4.211,\qquad
C_{10}=-4.103, $$ 
together with 
$$C_1=-0.257,\quad C_2=1.009,\quad C_3=-0.005,\quad
C_4=-0.078,\quad C_5=0.000,\quad C_6=0.001 .$$
The coefficient entering the vector semileptonic amplitude is taken as $q^2$-dependent,
\begin{equation}
    C_9^{\rm eff}(q^2)=C_9(\mu_b)+Y(q^2,\mu_b),
\end{equation}
where $Y(q^2,\mu_b)$ denotes the standard perturbative contribution generated by the four-quark operators. We evaluate the corresponding loop functions using $\overline m_c(\overline m_c)=1.275\,{\rm GeV}$.
The remaining numerical inputs entering the decay rate are collected in Table~\ref{tab:input parameters}. The $z$-expanded form factors are evaluated using the same conformal mapping as in Eq.~\eqref{eq:zparam}, with $t_+=(m_B+m_K)^2$ and
$t_0=(m_{\Lambda_b}-m_\Lambda)^2$, consistently with the mapping used to determine the corresponding $z$-expansion coefficients. 
The form factor uncertainties are propagated to the differential branching fraction by Monte Carlo sampling. The LCSR covariance includes the correlated $50\%$ estimate of missing perturbative corrections described above, while the lattice tensor form factor uncertainties are propagated from their corresponding covariance. No additional perturbative uncertainty is introduced at the level of the decay rate. At each value of $q^2$, we quote the median of the resulting $d\mathcal B/dq^2$ distribution as the central prediction and the interval between the $16^{\rm th}$ and $84^{\rm th}$ percentiles.

We stress that the perturbative contribution $Y(q^2,\mu_b)$ included in $C_9^{\rm eff}(q^2)$ is distinct from genuinely nonlocal hadronic contributions associated with time-ordered products of the electromagnetic current and the effective weak Hamiltonian. These effects are not calculated in the present work and can modify both the normalization and $q^2$ dependence of the decay amplitude. Accordingly, Fig.~\ref{fig:dBdq2_1_6} should be regarded as an illustrative prediction based on the local form factors and perturbative short-distance contributions, rather than a complete SM prediction.

Fig.~\ref{fig:dBdq2_1_6} shows the resulting differential branching fraction together with the LHCb measurements. Within the sizable form-factor uncertainties, the comparison provides an indication of the phenomenological impact of the LCSR vector and axial-vector form factors. In view of the external tensor
input and the omission of nonlocal hadronic contributions, no precision comparison with the experimental spectrum is attempted here. A fully self-contained phenomenological analysis would require tensor form factors and nonlocal hadronic contributions within a common framework and is left for future work.

\section{Summary and Discussion}
\label{sec:summary}
In this work, we have derived light-cone sum rules for the $\Lambda_b \to \Lambda$ form factors induced by pseudoscalar, vector, and axial-vector $b\to s$ currents, using vacuum-to-$\Lambda$ correlation functions and three-quark $\Lambda$-baryon DAs up to twist-$6$.
The resulting sum rules are valid at leading order in the strong coupling and provide a systematic framework for studying heavy-to-light baryonic transitions at low momentum transfer, while radiative $\alpha_s$ corrections are left for future study.
A central element of the construction is the explicit inclusion of the negative-parity $\Lambda_b^*(1/2^-)$ state in the hadronic representation. Suitable linear combinations of the invariant amplitudes are used to eliminate the $\Lambda_b^*$ pole contribution from the resulting sum rules.

We have considered pseudoscalar, axial-vector, and Ioffe interpolating currents for the $\Lambda_b$ baryon. All three choices lead to numerically stable LCSRs in the adopted Borel and continuum-threshold windows. At the level of central values, the pseudoscalar current gives the closest agreement with the expected EOM and HQS relations, while larger deviations are obtained for the axial-vector and Ioffe currents. Once the uncertainty associated with the missing perturbative corrections is included, however, these differences the enlarged uncertainties substantially weaken the discriminating power of the EOM/HQS relations, and the present analysis does not allow a statistically significant current selection. 

For the pseudoscalar current, the twist decomposition shows that the numerical hierarchy of the individual contributions is not monotonic, while the twist-$6$ contribution is small for all form factors considered. The resulting vector and axial-vector form factors are transformed to the helicity basis and parametrised using a truncated $z$-expansion. We have also compared the LCSR results with lattice-QCD determinations and constructed a common $z$-parametrisation using low-$q^2$ LCSR and high-$q^2$ lattice inputs.  

As an illustrative phenomenological application, we have evaluated the low-$q^2$ differential branching fraction for $\Lambda_b\to\Lambda\ell^+\ell^-$. For this purpose, the vector and axial-vector form factors are taken from the pseudoscalar-current LCSRs, while the tensor form factors are taken from lattice QCD. The perturbative four-quark contribution is included through $C_9^{\rm eff}(q^2)$, whereas genuine nonlocal hadronic contributions are not calculated. The resulting branching fraction prediction should therefore be viewed as an illustration of the phenomenological impact of the LCSR form factors rather than as a complete SM prediction. A calculation of perturbative corrections to the present LCSRs, together with a determination of the tensor form factors and nonlocal hadronic contributions within a consistent framework, would provide a natural extension of the present analysis.

\section*{Acknowledgment}
D.M. is grateful to Siavash Neshatpour for useful discussions. \\
This research is funded in part by the National Research Agency (ANR) under project no. ANR-21-CE31-0002-01.

\clearpage
\appendix

\section{Invariant amplitudes}
\label{app:A}
In this appendix we collect the explicit expressions for the invariant amplitudes entering the hadronic representation of the correlation functions with vector and axial-vector weak transition currents. They are obtained by inserting complete sets of intermediate $\Lambda_b$ and $\Lambda_b^*$ states into the correlation functions and expressing the resulting matrix elements in terms of the transition form factors defined in Sec.~\ref{sec:corrl}. The formulas given below are used to construct the hadronic dispersion relations and to identify the structures entering the light-cone sum rules.

For the vector transition current, the hadronic representation of the correlation function $F_\mu ^{(l)}(p,q)$ can be written in terms of six independent Lorentz structures, whose coefficients are expressed through the vector form factors $f_i(q^2)$ and $\hat{f}_i(q^2)$ associated with the $\Lambda_b$ and $\Lambda_b^*$ intermediate states, respectively:
\begin{align}
F^{(l)}_\mu(p,q) &= \frac{\lambda_{\Lambda_b}^{(l)}
m_{\Lambda_b}}{m_{\Lambda_b}^2-(p-q)^2} \bigg [ 2 f_1(q^2)\,p_\mu
-2\frac{f_2(q^2)}{m_{\Lambda_b}} \,p_\mu \not \! q
+(m_{\Lambda_b}-m_\Lambda)
\bigg ( f_1(q^2)- \frac{  m_{\Lambda_b}+m_\Lambda }{ m_{\Lambda_b} } f_2(q^2) \bigg )\gamma_{\mu}
\nonumber\\
 &+ \bigg ( f_1(q^2)- \frac{m_{\Lambda_b}+m_\Lambda }{ m_{\Lambda_b} } f_2(q^2) \bigg )  \gamma_{\mu}  \not \! q
+  \bigg ( -2 f_1(q^2) + \frac{m_{\Lambda_b}+m_\Lambda}{m_{\Lambda_b} } (f_2(q^2)+f_3(q^2)) \bigg )q_{\mu}
\nonumber\\
&  +\frac{1}{m_{\Lambda_b} } (f_2(q^2)-f_3(q^2))q_{\mu} \not \! q
\bigg ]  u_\Lambda({p})  \nonumber + \frac{\lambda_{\Lambda_b^*}^{(l)} m_{\Lambda_b^*}}{m_{\Lambda_b^*}^2-(p-q)^2}
\bigg [- 2\hat{f}_1(q^2)\,{p_\mu}
+2\frac{\hat{f}_2(q^2)}{m_{\Lambda_b^*}} \,{p_\mu} \not \! q
\nonumber \\
&+ (m_{\Lambda_b^*}+m_\Lambda) \bigg ( \hat{f}_1(q^2)+
\frac{m_{\Lambda_b^*}-m_\Lambda}{m_{\Lambda_b^*} } \hat{f}_2(q^2)
\bigg )\gamma_{\mu}
- \bigg ( \hat{f}_1(q^2)+\frac{m_{\Lambda_b^*}-m_\Lambda }{ {m_{\Lambda_b^*}}}
\hat{f}_2(q^2) \bigg )  \gamma_{\mu}  \not \! q
\nonumber\\
&+  \bigg ( 2 \hat{f}_1(q^2) + \frac{  m_{\Lambda_b^*}-m_\Lambda}{
m_{\Lambda_b^*} } (\hat{f}_2(q^2)+\hat{f}_3(q^2)) \bigg
)q_{\mu}
  -\frac{1}{ m_{\Lambda_b^*} }
(\hat{f}_2(q^2)-\hat{f}_3(q^2))q_{\mu} \not \! q
\bigg ]  u_\Lambda({p})    \nonumber \\
 &+ \int\limits_{s_0^h}^{\infty}\frac{ds}{s-(p-q)^2}
\bigg(\tilde{\rho}_{1}^{(l)}(s, q^2)\,p_\mu\,
+\tilde{\rho}_2^{(l)}\,p_\mu\slashed{q}
+\tilde{\rho}_3^{(l)}\,\gamma_\mu\,
+\tilde{\rho}_4^{(l)}\,\gamma_\mu\slashed{q}\,
+\tilde{\rho}_5^{(l)}\,q_\mu\,
+\tilde{\rho}_6^{(l)}\,q_\mu\slashed{q}\,\bigg)u_\Lambda({p})\,.
\label{eq:Fvec}
\end{align}
In Eq.~\eqref{eq:Fvec}, the first term corresponds to the ground-state $\Lambda_b$ pole, while the second term arises from the negative-parity excited state $\Lambda_b^*$. The couplings $\lambda_{\Lambda_b}^{(l)}$ and $\lambda_{\Lambda_b^*}^{(l)}$ are defined in Eqs.~\eqref{eq:decayc} and \eqref{eq:dclams}, respectively, while  $\tilde{\rho}_{i}^{(l)}(s, q^2)$ denotes the higher-state and continuum spectral density in the corresponding Lorentz structure.

For the axial-vector transition current, the corresponding hadronic representation reads
{\allowdisplaybreaks
\begin{align}
F^{(l)}_{\mu 5}({p},q) &= \frac{\lambda_{\Lambda_b}^{(l)}
m_{\Lambda_b}}{m_{\Lambda_b}^2-(p-q)^2} \bigg [ 2 g_1(q^2)\,p_\mu
-2\frac{g_2(q^2)}{m_{\Lambda_b}} \,p_\mu \not \! q
+(m_{\Lambda_b}-m_\Lambda)
\bigg ( g_1(q^2)- \frac{  m_{\Lambda_b}+m_\Lambda }{ m_{\Lambda_b} } g_2(q^2) \bigg )\gamma_{\mu}
\nonumber\\
 &+ \bigg ( g_1(q^2)- \frac{m_{\Lambda_b}+m_\Lambda }{ m_{\Lambda_b} } g_2(q^2) \bigg )  \gamma_{\mu}  \not \! q
+  \bigg ( -2 g_1(q^2) + \frac{m_{\Lambda_b}+m_\Lambda}{m_{\Lambda_b} } (g_2(q^2)+g_3(q^2)) \bigg )q_{\mu}
\nonumber\\
&  +\frac{1}{m_{\Lambda_b} } (g_2(q^2)-g_3(q^2))q_{\mu} \not \! q
\bigg ] \gamma_5 u_\Lambda({p})  \nonumber + \frac{\lambda_{\Lambda_b^*}^{(l)} m_{\Lambda_b^*}}{m_{\Lambda_b^*}^2-(p-q)^2}
\bigg [- 2\hat{g}_1(q^2)\,p_\mu
+2\frac{\hat{g}_2(q^2)}{m_{\Lambda_b^*}} \,p_\mu \not \! q
\nonumber \\
&+ (m_{\Lambda_b^*}+m_\Lambda) \bigg ( \hat{g}_1(q^2)+
\frac{m_{\Lambda_b^*}-m_\Lambda}{m_{\Lambda_b^*} } \hat{g}_2(q^2)
\bigg )\gamma_{\mu}
- \bigg ( \hat{g}_1(q^2)+\frac{m_{\Lambda_b^*}-m_\Lambda }{ {m_{\Lambda_b^*}}}
\hat{g}_2(q^2) \bigg )  \gamma_{\mu}  \not \! q
\nonumber\\
&+  \bigg ( 2 \hat{g}_1(q^2) + \frac{  m_{\Lambda_b^*}-m_\Lambda}{
m_{\Lambda_b^*} } (\hat{g}_2(q^2)+\hat{g}_3(q^2)) \bigg
)q_{\mu}
  -\frac{1}{ m_{\Lambda_b^*} }
(\hat{g}_2(q^2)-\hat{g}_3(q^2))q_{\mu} \not \! q
\bigg ]  \gamma_5 u_\Lambda({p})    \nonumber \\
 &+ \int\limits_{s_0^h}^{\infty}\frac{ds}{s-(p-q)^2}
\bigg(\bar{\rho}_{1}^{(l)}(s, q^2)\,p_\mu\,
+\bar{\rho}_2^{(l)}\,p_\mu\slashed{q}
+\bar{\rho}_3^{(l)}\,\gamma_\mu\,
+\bar{\rho}_4^{(l)}\,\gamma_\mu\slashed{q}\,
+\bar{\rho}_5^{(l)}\,q_\mu\,
+\bar{\rho}_6^{(l)}\,q_\mu\slashed{q}\,\bigg) \gamma_5 u_\Lambda({p})\,.
\label{eq:Faxvec}
\end{align}}%
Here, $g_i(q^2)$ and $\hat{g}_i(q^2)$ denote the axial-vector form factors for the $\Lambda_b$ and $\Lambda_b^*$ transitions, respectively. The Lorentz decomposition is chosen such that the coefficients of the Dirac structures can be directly matched to the invariant amplitudes introduced in Sec.~\ref{sec:corrl}. As in the vector case, the last term represents the contribution of higher-mass states and the continuum, encoded in the corresponding spectral densities.
In Eqs.~\eqref{eq:Fvec} and \eqref{eq:Faxvec}, the denominators multiplying $\hat f_2$ and $\hat g_2$ in the $\gamma_\mu\slashed q$ structures are $m_{\Lambda_b^*}$. This follows from the definitions of the negative-parity transition form factors, in which the corresponding Pauli-type terms are normalized by the mass of the negative-parity state, and is consistent with the other $\hat f_{2,3}$ and $\hat g_{2,3}$ terms in the same pole contributions.

\section{Light Cone Distribution Amplitudes }
\label{app:LCDAs}
In this appendix we summarise the definitions and conventions for the three-quark LCDAs of the $\Lambda$ baryon used in the light-cone OPE. They are defined through vacuum-to-$\Lambda$ matrix elements of nonlocal three-quark operators near the light-cone. We follow the conventions of~\cite{Liu:2008yg,Aliev:2010uy}: 
\begin{align}
 4 \bra{0} & \epsilon^{ijk}
u_\alpha^i(a_1 z) s_\beta^j(a_2 z) d_\gamma^k(a_3 z) \ket{\Lambda({p})} 
= {\cal S}_1 m_\Lambda C_{\alpha \beta} \left(\gamma_5 u_\Lambda
\right)_\gamma + {\cal S}_2 {m_\Lambda^2} C_{\alpha \beta}
\left(\!\not\!{z} \gamma_5 u_\Lambda \right)_\gamma + {\cal P}_1 m_\Lambda
\left(\gamma_5 C\right)_{\alpha \beta} {(u_\Lambda)}_\gamma \nonumber \\
& + {\cal P}_2 m_\Lambda^2 \left(\gamma_5 C \right)_{\alpha \beta}
\left(\!\not\!{z} u_\Lambda \right)_\gamma
 + \mathcal{V}_1
\left({\!\not\!{p}} C \right)_{\alpha \beta} \left(\gamma_5 u_\Lambda
\right)_\gamma + {\cal V}_2 m_\Lambda \left({\!\not\!{p}} C \right)_{\alpha \beta}
\left(\!\not\!{z} \gamma_5 u_\Lambda \right)_\gamma  + {\cal V}_3 m_\Lambda
\left(\gamma_\mu C \right)_{\alpha \beta}\left(\gamma^{\mu}
\gamma_5 u_\Lambda \right)_\gamma
 \nonumber \\
&+ {\cal V}_4 m_\Lambda^2 \left(\!\not\!{z}C \right)_{\alpha \beta}
\left(\gamma_5 u_\Lambda \right)_\gamma + {\cal V}_5 m_\Lambda^2
\left(\gamma_\mu C \right)_{\alpha \beta} \left(i \sigma^{\mu\nu}
z_\nu \gamma_5 u_\Lambda \right)_\gamma + {\cal V}_6 m_\Lambda^3 \left(\!\not\!{z} C \right)_{\alpha \beta}
\left(\!\not\!{z} \gamma_5 u_\Lambda \right)_\gamma  \nonumber \\
& +
\mathcal{A}_1
\left({\!\not\!{p}}\gamma_5 C \right)_{\alpha \beta} {(u_\Lambda)}_\gamma
+ {\cal A}_2 m_\Lambda \left({\!\not\!{p}}\gamma_5 C \right)_{\alpha
\beta} \left(\!\not\!{z} u_\Lambda \right)_\gamma  + {\cal A}_3 m_\Lambda
\left(\gamma_\mu \gamma_5 C \right)_{\alpha \beta}\left(
\gamma^{\mu} u_\Lambda \right)_\gamma \nonumber \\
& 
 + {\cal A}_4 m_\Lambda^2 \left(\!\not\!{z} \gamma_5 C \right)_{\alpha
\beta} {(u_\Lambda)}_\gamma + {\cal A}_5 m_\Lambda^2 \left(\gamma_\mu \gamma_5
C \right)_{\alpha \beta} \left(i \sigma^{\mu\nu} z_\nu u_\Lambda
\right)_\gamma  + {\cal A}_6 m_\Lambda^3 \left(\!\not\!{z} \gamma_5 C \right)_{\alpha
\beta} \left(\!\not\!{z} u_\Lambda \right)_\gamma \nonumber \\
&  +
\mathcal{T}_1
\left({p^\nu} i \sigma_{\mu\nu} C\right)_{\alpha \beta}
\left(\gamma^\mu\gamma_5 u_\Lambda \right)_\gamma  + {\cal T}_2 m_\Lambda \left(z^\mu {p^\nu} i \sigma_{\mu\nu}
C\right)_{\alpha \beta} \left(\gamma_5 u_\Lambda \right)_\gamma + {\cal
T}_3 m_\Lambda \left(\sigma_{\mu\nu} C\right)_{\alpha \beta}
\left(\sigma^{\mu\nu}\gamma_5 u_\Lambda \right)_\gamma  \nonumber \\
&+ {\cal T}_4 m_\Lambda \left({p^\nu} \sigma_{\mu\nu} C\right)_{\alpha
\beta} \left(\sigma^{\mu\rho} z_\rho \gamma_5 u_\Lambda \right)_\gamma +
{\cal T}_5 m_\Lambda^2 \left(z^\nu i \sigma_{\mu\nu} C\right)_{\alpha
\beta} \left(\gamma^\mu\gamma_5 u_\Lambda \right)_\gamma  + {\cal T}_6 m_\Lambda^2 \left(z^\mu {p^\nu} i \sigma_{\mu\nu}
C\right)_{\alpha \beta} \left(\!\not\!{z} \gamma_5 u_\Lambda
\right)_\gamma \nonumber \\
&+ {\cal T}_{7} m_\Lambda^2 \left(\sigma_{\mu\nu}
C\right)_{\alpha \beta} \left(\sigma^{\mu\nu} \!\not\!{z} \gamma_5
u_\Lambda \right)_\gamma  + {\cal T}_{8} m_\Lambda^3 \left(z^\nu \sigma_{\mu\nu}
C\right)_{\alpha \beta} \left(\sigma^{\mu\rho} z_\rho \gamma_5 u_\Lambda
\right)_\gamma\;.
\label{eq:lambdaDAs}
\end{align}
Here $z^\mu$ is a light-like vector $(z^2 = 0)$, $C$ denotes the charge-conjugation matrix, and $\alpha, \beta, \gamma$ are Dirac indices. The invariant functions $\mathcal{S}_{1,2}$, $\mathcal{P}_{1,2}$,
$\mathcal{V}_{1,...,6}$, $\mathcal{A}_{1,...,6}$,
$\mathcal{T}_{1,...,8}$ depend on the longitudinal momentum fractions of the constituent quarks and encode the nonperturbative structure of the $\Lambda$ baryon. The invariant functions are represented in momentum-fraction space as
\begin{equation}
\mathcal{M}(a_1,a_2,a_3,(p\cdot z))= \int \! d x_1 d x_2 d x_3
 \delta(1\!-\!x_1\!-\!x_2\!-\!x_3)e^{-i
(p  \cdot  z) \sum_i x_i a_i} M(x_i).
\label{eq:F1}
\end{equation}
The variables $x_i$ denote the longitudinal momentum fractions carried by the three quarks and satisfy {inside the $\Lambda$ baryon, satisfying} $x_1 + x_2 +x_3 = 1$. In the present calculation we choose the light-cone configuration $a_1 = a_3 = 0,\, a_2 =1,$ corresponding to the strange quark located at position $z$. The relations between the invariant functions in \eqref{eq:lambdaDAs} and the corresponding momentum-space functions entering \eqref{eq:F1} are collected in Table \ref{table:calF_F}. 
\begin{table}[h]
\begin{center}
\begin{tabular}{|c|c||c|c|}
\hline
\hline
 $\mathcal{M}$  & integrand on r.h.s. of (\ref{eq:F1})&  $\mathcal{M}$   & integrand on r.h.s. of (\ref{eq:F1}) \\
\hline ${\cal S}_1$ & $S_1$    & $2 (p \! \cdot \! z) \, {\cal
S}_2$ &
$S_1-S_2 $\\
\hline
 ${\cal P}_1$ & $P_1$ & $2 (p \! \cdot \! z) \, {\cal P}_2$ &  $P_2-P_1 $ \\
 \hline
 ${\cal V}_1$ & $ V_1$ & $2 (p \! \cdot
\! z) \, {\cal V}_2 $ &  $V_1 - V_2 - V_3$\\
 \hline
 $2 {\cal V}_3$ & $V_3$ & $4 (p \! \cdot \! z)
\, {\cal V}_4$ & $- 2 V_1 + V_3 + V_4  + 2 V_5$ \\
 \hline
 $4 (p \! \cdot \! z) {\cal V}_5$ & $V_4 - V_3$ & $4 \left(p \! \cdot \! z\right)^2  {\cal V}_6$ &  $- V_1 + V_2 +
V_3 +  V_4 + V_5 - V_6$ \\
 \hline
 $ {\cal A}_1$ & $A_1$ & $2 (p \! \cdot \! z) {\cal A}_2 $ &  $- A_1 + A_2 -  A_3$ \\
 \hline
 $ 2 {\cal A}_3$ & $A_3$ & $4 (p \! \cdot \! z){\cal A}_4 $&  $- 2 A_1 - A_3 - A_4  + 2 A_5$ \\
 \hline
 $4 (p \! \cdot \! z) {\cal A}_5$ & $A_3 - A_4$ & $4 \left(p \! \cdot \! z\right)^2  {\cal A}_6$ & $A_1 - A_2 +  A_3 +  A_4 - A_5 + A_6$ \\
 \hline
 ${\cal T}_1 $ & $T_1$ &  $ 2 (p \! \cdot \! z) {\cal T}_2$ &  $T_1 + T_2 - 2 T_3$\\
 \hline
  $2 {\cal T}_3$& $T_7$ & $2 (p \! \cdot \! z){\cal T}_4 $ &  $T_1 - T_2 - 2  T_7$ \\
 \hline
 $ 2 (p \! \cdot \! z) {\cal T}_5$ & $- T_1 + T_5 + 2  T_8$ & $4 \left(p \! \cdot \! z\right)^2 {\cal T}_6$ & $ 2 T_2 - 2 T_3 - 2 T_4 + 2 T_5 + 2 T_7 + 2 T_8$ \\
 \hline
 $ 4 (p \! \cdot \! z) {\cal T}_7 $ & $ T_7 - T_8$ & $4 \left(p \! \cdot \! z\right)^2 {\cal T}_8$ &  $-T_1 + T_2 + T_5 - T_6 + 2 T_7 + 2 T_8 $ \\
 \hline
 \hline
\end{tabular}
\end{center}
\caption{ Relations between the invariant functions appearing in \eqref{eq:lambdaDAs} and the corresponding momentum-space functions defined in}~\eqref{eq:F1}. 
\label{table:calF_F}
\end{table}

The explicit expressions for the $\Lambda$-baryon LCDAs up to twist-6 used in this work are listed below.

Twist-3 DAs:
\begin{align}
    V_1 =& 0\,, \hspace{2.cm}
    A_1 = -120\, x_1 x_2 x_3\, \phi_3^0\,, \hspace{2.cm}
    T_1  = 0 \,.
\end{align}

Twist-4 DAs:
\begin{align}
    S_1  =& 6\,x_3(1-x_3) (\xi_4^0 + \xi_4^{0\,'})\,, \hspace{1cm}
    P_1 = 6\,(1-x_3) (\xi_4^0 - \xi_4^{0\,'})\,, \hspace{1cm}
    V_2  =  0\,, \nonumber \\
    V_3 =& 12\,(x_1 - x_2) x_3 \,\psi_4^0\,, \hspace{1.8cm}
    A_2 = -24\,x_1 x_2 \,\phi_4^0 \,,\hspace{2.2cm}
    A_3 = -12\, x_3(1 - x_3)  \,\psi_4^0\,, \nonumber \\
    T_2 =& 0, \hspace{2cm} T_3 =  6\,x_3(x_2-x_1) (-\xi_4^0 + \xi_4^{0\,'})\,,\hspace{1.5cm}
    T_7  = - 6\,x_3(x_1-x_2) (\xi_4^0 + \xi_4^{0\,'})\,. 
\end{align}

Twist-5 DAs:
\begin{align}
    S_2 & = \frac{3}{2}\, (x_1 + x_2) (\xi_5^0 + \xi_5^{0\,'})\,, \qquad
    P_2 = \frac{3}{2}\,(x_1 + x_2) (\xi_5^0 - \xi_5^{0\,'})\,, \qquad
    V_4  = 3\,(x_2 - x_1) \,\psi_5^0\, , \nonumber \\
    V_5 &= 0\,, \qquad
    A_4 = -3\,(1 - x_3 )\,\psi_5^0\,, \qquad
    A_5 = -6\, x_3  \,\phi_5^0\,, \qquad
    T_4 = - \frac{3}{2}\, (x_1 - x_2) (\xi_5^0 + \xi_5^{0\,'})\,, \nonumber \\
    T_5 &=  0\,, \qquad
    T_8 = - \frac{3}{2}\, (x_1 - x_2) (\xi_5^0 - \xi_5^{0\,'})\,. 
\end{align}

Twist-6 DAs:
\begin{align}
   V_6 &= 0\,, \hspace{2.cm}
    A_6 = -2\, \phi_6^0\,, \hspace{2.cm}
    T_6  = 0\,.  
\end{align}
The lowest conformal coefficients entering these DAs are related to the nonperturbative parameters $f_\Lambda$, $\lambda_1$, $\lambda_2$, and $\lambda_3$ by 
\begin{align}
    \phi_{3}^{0} & = \phi_6^0 = -f_\Lambda, \hspace{2 cm} \phi_4^0 =\phi_5^0 = -\frac{1}{2} (f_\Lambda + \lambda_1)   \nonumber \\ 
    \psi_{4}^{0} & = \psi_5^0 = -\frac{1}{2} (f_\Lambda - \lambda_1) , \hspace{1.5 cm} \xi_4^0 =\xi_5^0 =  (\lambda_2 + \lambda_3), \hspace{1.5 cm}  \xi_4^{0'} =\xi_5^{0'} =  (\lambda_3 - \lambda_2).
\end{align}
For compactness, we use the following combinations of $\Lambda$-baryon DAs:

\begin{align}
   S_{12} =& S_1 - S_2\,,       &P_{21} = P_2 - P_1 \,,
\nonumber\\[6pt]
 V_{1345} =& -2V_1+V_3+V_4+2 V_5\,,     &V_{43} = V_4-V_3\,, \nonumber \\
 V_{123456} =&
-V_1+V_2+V_3+V_4+V_5-V_6\,,   &V_{123} = V_1-V_2-V_3
\nonumber\\[6pt]
  A_{1345} =& -2A_1-A_3-A_4+2 A_5\,,     &A_{34} = A_3-A_4 \nonumber \\
A_{123456} =& A_1-A_2+A_3+A_4-A_5+A_6\,,   & A_{123} =
-A_1+A_2-A_3
\nonumber\\[6pt]
 T_{78}   =& T_7 - T_8\,,     & T_{123}   = T_1 + T_2 - 2T_3
\,,
\nonumber\\
 T_{234578}   = &2 T_2 - 2 T_3 - 2 T_4 + 2 T_5 + 2 T_7 + 2 T_8 \,,
  &T_{127}   = T_1 - T_2 - 2 T_7\,, \nonumber \\
 T_{125678}   =& -T_1 + T_2 + T_5 - T_6 + 2
T_7 + 2 T_8 \,,   &T_{158}   = -T_1 + T_5 + 2 T_8\,.
\end{align}
These DAs enter the light-cone sum rules through convolution integrals with the perturbatively calculable hard-scattering kernels, as described in Sec.~\ref{sec:lcope}.

We further introduce the integrated DAs
\begin{align}
    \tilde{M} (x_2) =& \int_0^{1-x_2}  dx_1 M(x_1, x_2, 1 - x_1 - x_2)\,, \nonumber \\ 
    \tilde{\tilde{M}} (x_2) =& \int_1^{x_2} dx_2^{'} \int_0^{1-x_2^{'}}  dx_1 M(x_1, x_2^{'}, 1 - x_1 - x_2^{'})\,, \nonumber \\ 
    \tilde{\tilde{\tilde{M}}} (x_2) =& \int_1^{x_2} dx_2^{'} \int_1^{x_2^{'}} dx_2^{''}\int_0^{1-x_2^{''}}  dx_1 M(x_1, x_2^{''}, 1 - x_1 - x_2^{''}). 
    \label{eq:Mtildes}
\end{align}
These definitions arise when partial integrations in $x_2$ are used to remove explicit factors of $1/p\cdot z$ from the light-cone expressions.  

\section{QCD representation of correlation functions}
\label{app:qcd-representation}
In this appendix we collect the QCD representations of the invariant amplitudes entering the light-cone sum rules for the $\Lambda_b \to \Lambda$ form factors. The expressions follow from the light-cone OPE described in Sec.~\ref{sec:lcope} and are written in terms of convolution integrals of hard-scattering kernels with $\Lambda$-baryon LCDAs.

\subsection{Pseudoscalar transition current}
\label{subapp:pseudo}
For the pseudoscalar weak transition current $j_5 = m_b \bar{s} i \gamma_5 b$, {the invariant amplitudes $F^{(l)}_j((p-q)^2,q^2)$ are written in the form of Eq.~(\ref{eq:qcdcor}) with coefficient functions $T^{(l)}_{jn}$.} The expressions below are given for the pseudoscalar ($\mathcal{P}$) and axial-vector ($\mathcal{A}$) interpolating currents of the $\Lambda_b$ baryon. For the Ioffe current, the corresponding expressions are considerably lengthier and are implemented directly in the Mathematica notebook used for the numerical analysis.
\begin{itemize}
\item The pseudoscalar interpolating current:
\begin{eqnarray}
 T^{(\cal P)}_{11} &=& \frac{m_\Lambda}{2} \Big[ (m_b-x\,m_\Lambda) \Phi_1^{(\cal P)} - m_\Lambda \Phi_2^{(\cal P)} \Big] \,,\nonumber\\
 T^{(\cal P)}_{12}  &=& -\frac{m_\Lambda^2}{2} \Big[m_b\Big(m_b-x \,m_\Lambda\Big) \Phi_2^{(\cal P)} + 2x m_\Lambda^2 \Phi_3^{(\cal P)} \Big]\,, \nonumber\\
T^{(\cal P)}_{13}   &=& 2 m_\Lambda^3 m_b^2 \big(m_b-x \,m_\Lambda\big)
\Phi_3^{(\cal P)}\,, \nonumber
\end{eqnarray}

\begin{eqnarray}
T^{(\cal P)}_{21} &=& -\frac{m_\Lambda}{2} \Phi_1^{(\cal P)} \,,
\qquad
 T^{(\cal P)}_{22} = \frac{m_\Lambda^2}{2}\Big( m_b \Phi_2^{(\cal P)} -
2 m_\Lambda \Phi_3^{(\cal P)} \Big)\,,
\nonumber\\
 T^{(\cal P)}_{23} &=& - 2 m_\Lambda^3 m_b^2 \Phi_3^{(\cal P)} \,,
\end{eqnarray}
where the functions $\Phi_i^{(\cal P)}$ in the above equations are:
\begin{eqnarray}
 \Phi_1^{(\cal P)} &=& 2 \tilde{A}_1+4 \tilde{A}_3+2 \tilde{A}_{123}+2 \tilde{P}_1+2 \tilde{S}_1
 +6 \tilde{T}_1-12 \tilde{T}_7-\tilde{T}_{123}-5 \tilde{T}_{127}\nonumber \\
   &&-2 \tilde{V}_1+4 \tilde{V}_3+2
   \tilde{V}_{123} \,, \nonumber\\
 \Phi_2^{(\cal P)} &=& 3 \tilde{\tilde{A}}_{34}+2 \tilde{\tilde{A}}_{123}-\tilde{\tilde{A}}_{1345}
 -2 \tilde{\tilde{P}}_{21}+2 \tilde{\tilde{S}}_{12}-12 \tilde{\tilde{T}}_{78}-2
   \tilde{\tilde{T}}_{123}-4 \tilde{\tilde{T}}_{127}\nonumber \\
   &&-6 \tilde{\tilde{T}}_{158}
   +\tilde{\tilde{T}}_{234578} -3 \tilde{\tilde{V}}_{43}+2
   \tilde{\tilde{V}}_{123}+\tilde{\tilde{V}}_{1345} \,, \nonumber\\
 \Phi_3^{(\cal P)} &=& \tilde{\tilde{\tilde{A}}}_{123456}-3
   \tilde{\tilde{\tilde{T}}}_{125678}+\tilde{\tilde{\tilde{T}}}_{234578}+\tilde{\tilde{\tilde{V}}}_{123456}
   \,. \nonumber
\end{eqnarray}
Here the shorthand notations $\tilde{M},\, \tilde{\tilde{M}}$, and $\tilde{\tilde{\tilde{M}}}$ (where, $M \in {\{S,P,V,A,T\}}$) denote the integrated $\Lambda$-baryon
DAs defined in Eq.~\eqref{eq:Mtildes}.

\item The axial-vector interpolating current:
\begin{eqnarray}
T^{(\cal A)}_{11}&=& 2{ m_b^2 -q^2 \over x} \Phi_1^{(\cal A)}
+ x \,m_\Lambda^2 \Big[2\Phi_1^{(\cal A)}+\Phi_2^{(\cal
A)}\Big] + m_\Lambda m_b \Phi_3^{(\cal A)}  + m_\Lambda^2\Big[\Phi_4^{(\cal A)}+\frac{2\Phi_5^{(\cal A)}}{x}\Big] \,, \nonumber\\
 T^{(\cal A)}_{12}&=& -m_\Lambda^2 \bigg[2(x^2\,m_\Lambda^2-q^2) \Phi_6^{(\cal A)} +
x\,m_\Lambda m_b \Phi_7^{(\cal A)} + m_b^2 \Phi_8^{(\cal A)} -
2\frac{q^2+m_b^2}{x} \Phi_5^{(\cal A)}\nonumber \\
&& + 2 x \,m_\Lambda^2 \Phi_9^{(\cal A)} \bigg] \,,  \nonumber\\
T^{(\cal A)}_{13} &=& 4 \frac{m_\Lambda^2}{x} \bigg[ m_b^2
(q^2-m_b^2) \Phi_5^{(\cal A)} - x^2 m_\Lambda^2 m_b^2 \Phi_9 ^{(\cal
A)}+ x m_\Lambda m_b^3 \Phi_{10}^{(\cal A)} \bigg]\,, \nonumber
\end{eqnarray}

\begin{eqnarray}
 T^{(\cal A)}_{21} &=& \frac{m_\Lambda}{x} \bigg[2\Phi_6^{(\cal A)} + x \,\Phi_2^{(\cal A)} \bigg] \,, \qquad
 T^{(\cal A)}_{23} =  4 m_\Lambda^3 m_b^2 \Phi_{11}^{(\cal A)}\,,  \nonumber\\
 T^{(\cal A)}_{22}&=& \frac{m_\Lambda}{x} \bigg[2(q^2-m_b^2-x^2\,m_\Lambda^2) \Phi_6^{(\cal A)}
 - x \,m_\Lambda m_b \Phi_7^{(\cal A)} + 2 x\,m_\Lambda^2 \Phi_{11}^{(\cal A)} \bigg]
 \,, \,\,\,\,\,\,
\end{eqnarray}
where the functions $\Phi_i^{(\cal A)}$ are:
{\allowdisplaybreaks
\begin{eqnarray}
 \Phi_1^{(\cal A)} &=& \tilde{A}_1+2 \tilde{T}_1+\tilde{V}_1 \,, \nonumber\\
 \Phi_2^{(\cal A)} &=& 2 \tilde{A}_3-2 \tilde{P}_1+2 \tilde{S}_1-2 \tilde{T}_1+\tilde{T}_{123}+\tilde{T}_{127}-2 \tilde{V}_3  \,, \nonumber\\
 \Phi_3^{(\cal A)} &=& 2 \tilde{A}_1+4 \tilde{A}_3+2 \tilde{A}_{123}-4 \tilde{P}_1+4 \tilde{S}_1+2 \tilde{V}_1-4 \tilde{V}_3-2
   \tilde{V}_{123}  \,, \nonumber\\
 \Phi_4^{(\cal A)} &=& -2 \tilde{\tilde{A}}_{1345}+2 \tilde{\tilde{P}}_{21}+2 \tilde{\tilde{S}}_{12}-2 \tilde{\tilde{T}}_{123}
 +4 \tilde{\tilde{T}}_{127}-6 \tilde{\tilde{T}}_{158}+3 \tilde{\tilde{T}}_{234578}-2 \tilde{\tilde{V}}_{1345}  \,, \nonumber\\
 \Phi_5^{(\cal A)} &=& -\tilde{\tilde{\tilde{T}}}_{234578}  \,, \nonumber\\
 \Phi_6^{(\cal A)} &=& \tilde{\tilde{A}}_{123}-\tilde{\tilde{T}}_{123}-\tilde{\tilde{T}}_{127}-\tilde{\tilde{V}}_{123}  \,, \nonumber\\
 \Phi_7^{(\cal A)} &=& -3 \tilde{\tilde{A}}_{34}-2 \tilde{\tilde{A}}_{123}+\tilde{\tilde{A}}_{1345}-4 \tilde{\tilde{P}}_{21}
 -4 \tilde{\tilde{S}}_{12}-3 \tilde{\tilde{V}}_{43}+2 \tilde{\tilde{V}}_{123}+\tilde{\tilde{V}}_{1345}  \,, \nonumber\\
 \Phi_8^{(\cal A)} &=& 2 \tilde{\tilde{A}}_{123}+2 \tilde{\tilde{A}}_{1345}-2 \tilde{\tilde{P}}_{21}-2 \tilde{\tilde{S}}_{12}
 -6 \tilde{\tilde{T}}_{127}+6 \tilde{\tilde{T}}_{158}-3
 \tilde{\tilde{T}}_{234578}-2 \tilde{\tilde{V}}_{123}+2 \tilde{\tilde{V}}_{1345}  \,, \nonumber \\
 \Phi_9^{(\cal A)} &=& -2 \tilde{\tilde{\tilde{A}}}_{123456}
 +3 \tilde{\tilde{\tilde{T}}}_{125678}-2 \tilde{\tilde{\tilde{T}}}_{234578}+2 \tilde{\tilde{\tilde{V}}}_{123456}  \,, \nonumber\\
 \Phi_{10}^{(\cal A)} &=& \tilde{\tilde{\tilde{A}}}_{123456}-\tilde{\tilde{\tilde{V}}}_{123456}  \,, \nonumber\\
 \Phi_{11}^{(\cal A)} &=& 2 \tilde{\tilde{\tilde{A}}}_{123456}-3 \tilde{\tilde{\tilde{T}}}_{125678}
 + \tilde{\tilde{\tilde{T}}}_{234578}-2\tilde{\tilde{\tilde{V}}}_{123456}
 \,. \nonumber
\end{eqnarray}
}
\end{itemize}

\subsection{Vector transition current}
\label{app:vector-current}
For the vector weak transition current $j_\mu = \bar{s} \gamma_\mu b$, the invariant amplitudes ${\tilde{F}^{(l)}_j}((p-q)^2,q^2)$ for the correlation function with the vector transition current $j_{\mu}$ are given by Eq.~(\ref{eq:qcdcor}) with the
replacement of the coefficient $m_b/4 \to 1/4$. The coefficient functions ${\tilde{T}^{(l)}_{jn}}$ with ${l}={\cal P},{\cal A}$, $j=1,2,...6$ and $n=1,2,3$ are listed below for:

\begin{itemize}
\item{The pseudoscalar interpolating current:}
\begin{eqnarray}
\tilde{T}^{(\cal P)}_{11} &=& x \,m_\Lambda \tilde{\Phi}_1^{(\cal
P)}  \, , \qquad \tilde{T}^{(\cal P)}_{12} = x \,m_\Lambda^3
\bigg[x\,\tilde{\Phi}_2^{(\cal P)}  + 2 \tilde{\Phi}_3^{(\cal
P)}  \bigg] \,, \nonumber\\
\tilde{T}^{(\cal P)}_{13} &=& 4 x \,m_\Lambda^3 m_b^2
\tilde{\Phi}_3^{(\cal P)}  \,, \nonumber
\end{eqnarray}
\begin{eqnarray}
\tilde{T}^{(\cal P)}_{21} &=&  \tilde{T}^{(\cal
P)}_{23}= 0 \,, \qquad   \tilde{T}^{(\cal P)}_{22} = - x
\,m_\Lambda^2 \tilde{\Phi}_2^{(\cal P)}  \,, \nonumber
\end{eqnarray}
\begin{eqnarray}
\tilde{T}^{(\cal P)}_{31}&=& \frac{m_\Lambda}{2}(m_b-x \,m_\Lambda)
\tilde{\Phi}_1^{(\cal
P)}  \,, \nonumber\\
\tilde{T}^{(\cal P)}_{32} &=&
-\frac{m_\Lambda^2}{2}\bigg[m_b(m_b-x \,m_\Lambda) \tilde{\Phi}_2^{(\cal P)} +
2 x \,m_\Lambda^2 \tilde{\Phi}_3^{(\cal
P)} \bigg] \,, \nonumber\\
\tilde{T}^{(\cal P)}_{33}&=& 2 m_\Lambda^3 m_b^2 (m_b-x \,m_\Lambda)\,
\tilde{\Phi}_3^{(\cal P)} \,, \nonumber
\end{eqnarray}
\begin{eqnarray}
\tilde{T}^{(\cal P)}_{41}&=& \frac{m_\Lambda}{2}
\tilde{\Phi}_1^{(\cal P)} \,, \qquad  \tilde{T}^{(\cal
P)}_{42} =  \frac{m_\Lambda^2}{2} \Big[- m_b \tilde{\Phi}_2^{(\cal P)} +
2 m_\Lambda \tilde{\Phi}_3^{(\cal
P)} \Big] \,, \nonumber\\
\tilde{T}^{(\cal P)}_{43} &=& 2 m_\Lambda^3 m_b^2
\tilde{\Phi}_3^{(\cal P)} \,, \nonumber
\end{eqnarray}
\begin{eqnarray}
\tilde{T}^{(\cal P)}_{51}&=& - m_\Lambda \tilde{\Phi}_1^{(\cal P)}
\,, \qquad   \tilde{T}^{(\cal P)}_{52}= -m_\Lambda^3
\bigg[x\,\tilde{\Phi}_2^{(\cal P)} + 2 \tilde{\Phi}_3^{(\cal
P)} \bigg] \,, \nonumber\\
\tilde{T}^{(\cal P)}_{53}&=& - 4 m_\Lambda^3 m_b^2
\tilde{\Phi}_3^{(\cal P)} \,, \nonumber
\end{eqnarray}
\begin{eqnarray}
\tilde{T}^{(\cal P)}_{61}=\tilde{T}^{(\cal P)}_{63}= 0
\,, \qquad \tilde{T}^{(\cal P)}_{62}= m_\Lambda^2
\tilde{\Phi}_2^{(\cal P)} \,,
\end{eqnarray}
where the functions $\tilde{\Phi}_i^{(\cal P)}$ are given by:
\begin{eqnarray}
\tilde{\Phi}_1^{(\cal P)} &=& 2 \tilde{A}_1+4 \tilde{A}_3+2
\tilde{A}_{123}+2 \tilde{P}_1+2 \tilde{S}_1+6 \tilde{T}_1-12
\tilde{T}_7-\tilde{T}_{123}-5 \tilde{T}_{127}\nonumber
\\ && -2 \tilde{V}_1+4 \tilde{V}_3+2 \tilde{V}_{123} \,,\nonumber\\
\tilde{\Phi}_2^{(\cal P)}  &=& 3 \tilde{\tilde{A}}_{34}+2
\tilde{\tilde{A}}_{123}-\tilde{\tilde{A}}_{1345}-2
\tilde{\tilde{P}}_{21}+2 \tilde{\tilde{S}}_{12}-12
\tilde{\tilde{T}}_{78}-2 \tilde{\tilde{T}}_{123}-4
\tilde{\tilde{T}}_{127}\nonumber
\\ &&-6 \tilde{\tilde{T}}_{158}+\tilde{\tilde{T}}_{234578} -3 \tilde{\tilde{V}}_{43}+2
   \tilde{\tilde{V}}_{123}+\tilde{\tilde{V}}_{1345} \,,\nonumber\\
\tilde{\Phi}_3^{(\cal P)} &=&\tilde{\tilde{\tilde{A}}}_{123456}-3
   \tilde{\tilde{\tilde{T}}}_{125678}+\tilde{\tilde{\tilde{T}}}_{234578}+\tilde{\tilde{\tilde{V}}}_{123456}\,. \nonumber
\end{eqnarray}

\item{The axial-vector interpolating current:}
\begin{eqnarray}
\tilde{T}^{(\cal A)}_{11}  &=& 2\Big[2 m_b
\tilde{\Phi}_1^{(\cal A)} - x \,m_\Lambda (2\tilde{\Phi}_1^{(\cal A)}
+ \tilde{\Phi}_2^{(\cal A)}) + 2 m_\Lambda \tilde{\Phi}_3^{(\cal A)} \Big] \,, \nonumber\\
\tilde{T}^{(\cal A)}_{12}  &=& 2 m_\Lambda \Big[ x^2\,m_\Lambda^2
\tilde{\Phi}_4^{(\cal A)}
 + x\,m_\Lambda m_b \tilde{\Phi}_5^{(\cal A)} + 2 m_b^2 \tilde{\Phi}_3^{(\cal A)}
 + 2 x\,m_\Lambda^2 \tilde{\Phi}_6^{(\cal A)} \Big] \,, \nonumber\\
\tilde{T}^{(\cal A)}_{13}  &=& 8 m_\Lambda^2 m_b \Big[ m_b^2
\tilde{\Phi}_7^{(\cal A)} + x\,m_\Lambda m_b \tilde{\Phi}_6^{(\cal A)} +
x^2\,m_\Lambda^2 \tilde{\Phi}_8^{(\cal A)} \Big] \,, \nonumber
\end{eqnarray}
\begin{eqnarray}
\tilde{T}^{(\cal A)}_{21} &=& 4 \tilde{\Phi}_1^{(\cal A)}
\,,  \qquad \tilde{T}^{(\cal A)}_{23}  =  8 m_\Lambda^2 m_b \Big[
m_b \tilde{\Phi}_7^ {(\cal A)} - x \,m_\Lambda \tilde{\Phi}_8^{(\cal
A)} \Big] \,, \nonumber\nonumber\\
\tilde{T}^{(\cal A)}_{22}  &=& 2 m_\Lambda \Big[ 2 m_b
\tilde{\Phi}_3^{(\cal A)} - x \,m_\Lambda \tilde{\Phi}_4^{(\cal A)} + 2
m_\Lambda \tilde{\Phi}_7^{(\cal A)} \Big] \,, \nonumber
\end{eqnarray}
\begin{eqnarray}
\tilde{T}^{(\cal A)}_{31}&=& 2 { m_b^2 -q^2 \over x}
\tilde{\Phi}_1^{(\cal A)}
 + m_\Lambda m_b \tilde{\Phi}_2^{(\cal A)} + m_\Lambda^2 \tilde{\Phi}_{9}^{(\cal A)} -2 x \,m_\Lambda^2 \tilde{\Phi}_{10}^{(\cal A)}- 2 \frac{m_\Lambda^2}{x} \tilde{\Phi}_7^{(\cal A)}  \,, \nonumber\\
\tilde{T}^{(\cal A)}_{32} &=& m_\Lambda^2 \Big[
2(q^2-x^2\,m_\Lambda^2)\tilde{\Phi}_3^{(\cal A)} - 2
\frac{q^2+m_b^2}{x}\tilde{\Phi}_7^{(\cal A)}
  + x m_\Lambda m_b \tilde{\Phi}_{11}^{(\cal A)} \nonumber \, \\
  && + m_b^2 (\tilde{\Phi}_5^{(\cal A)}-2 \tilde{\Phi}_3^{(\cal A)}) + 2 m_\Lambda(m_b + x \,m_\Lambda) \tilde{\Phi}_8^{(\cal A)} \Big] \nonumber \,, \\
\tilde{T}^{(\cal A)}_{33}&=&  \frac{4 m_b^2 m_\Lambda^2}{x} \Big[
(m_b^2-q^2)\tilde{\Phi}_7^{(\cal A)} + x m_\Lambda m_b
\tilde{\Phi}_{12}^{(\cal A)} + x^2\,m_\Lambda^2 \tilde{\Phi}_8^{(\cal
A)} \Big] \,, \nonumber
\end{eqnarray}
\begin{eqnarray}
\tilde{T}^{(\cal A)}_{41} &=& 2 m_\Lambda\Big[
(\tilde{\Phi}_1^{(\cal A)}  + \tilde{\Phi}_{10}^{(\cal A)} ) -
\frac{\tilde{\Phi}_3^{(\cal A)} }{x} \Big] \,, \qquad \tilde{T}^{(\cal A)}_{43} = 4 m_\Lambda^3 m_b^2
\tilde{\Phi}_{13}^{(\cal A)} \,,  \nonumber\\
\tilde{T}^{(\cal A)}_{42} &=& \frac{m_\Lambda}{x} \Big[ 2
(m_b^2+x^2\,m_\Lambda^2-q^2) \tilde{\Phi}_3^{(\cal A)}  - x\,m_\Lambda m_b
\tilde{\Phi}_{11}^{(\cal A)}  +2 x\,m_\Lambda^2 \tilde{\Phi}_{13}^{(\cal
A)}  \Big]  \,, \nonumber
\end{eqnarray}
\begin{eqnarray}
\tilde{T}^{(\cal A)}_{51}  &=& 2 m_\Lambda \tilde{\Phi}_2^{(\cal
A)} \,, \qquad  \tilde{T}^{(\cal A)}_{53} = 8 m_\Lambda^3 m_b
\Big[ m_b \tilde{\Phi}_{14}^{(\cal A)} - x \,m_\Lambda
\tilde{\Phi}_8^{(\cal A)}
\Big] \,, \nonumber \\
\tilde{T}^{(\cal A)}_{52}  &=& 2 m_\Lambda^2 \Big[ -x \,m_\Lambda
\tilde{\Phi}_4^{(\cal A)} - m_b (\tilde{\Phi}_5^{(\cal
A)}+2\tilde{\Phi}_3^{(\cal A)}) + 2 m_\Lambda \tilde{\Phi}_{14}^{(\cal
A)}  \Big] \,, \nonumber
\end{eqnarray}
\begin{eqnarray}
\tilde{T}^{(\cal A)}_{61} &=& 0 \,, ~~    \tilde{T}^{(\cal A)}_{62}  = 2 m_\Lambda^2 \tilde{\Phi}_4^{(\cal A)} \,,
~~ \tilde{T}^{(\cal A)}_{63}  = 8 m_\Lambda^3 m_b
\tilde{\Phi}_8^{(\cal A)} \,.
\end{eqnarray}

The functions $\tilde{\Phi}_i^{(\cal A)}$ are given by:
\begin{eqnarray}
\tilde{\Phi}_1^{(\cal A)} &=& \tilde{A}_1+2 \tilde{T}_1+\tilde{V}_1 \,, \nonumber\\
\tilde{\Phi}_2^{(\cal A)} &=& 2 \tilde{A}_3-2 \tilde{P}_1+2 \tilde{S}_1-2 \tilde{T}_1
+\tilde{T}_{123}+\tilde{T}_{127}-2 \tilde{V}_3 \,, \nonumber\\
\tilde{\Phi}_3^{(\cal A)} &=& \tilde{\tilde{A}}_{123}-\tilde{\tilde{T}}_{123}
-\tilde{\tilde{T}}_{127}-\tilde{\tilde{V}}_{123} \,, \nonumber\\
\tilde{\Phi}_4^{(\cal A)} &=&
-\tilde{\tilde{A}}_{34}+\tilde{\tilde{A}}_{1345}-2
\tilde{\tilde{P}}_{21}-2 \tilde{\tilde{S}}_{12}-2
\tilde{\tilde{T}}_{127}+2\tilde{\tilde{T}}_{158}-\tilde{\tilde{T}}_{234578}
-\tilde{\tilde{V}}_{43}+\tilde{\tilde{V}}_{1345} \,, \nonumber\\
\tilde{\Phi}_5^{(\cal A)} &=& -\tilde{\tilde{A}}_{34}-2
\tilde{\tilde{A}}_{123}-\tilde{\tilde{A}}_{1345}+4
\tilde{\tilde{T}}_{127}-4 \tilde{\tilde{T}}_{158}+2
 \tilde{\tilde{T}}_{234578}-\tilde{\tilde{V}}_{43}+2 \tilde{\tilde{V}}_{123}-\tilde{\tilde{V}}_{1345} \,, \nonumber\\
\tilde{\Phi}_6^{(\cal A)} &=& - \tilde{\tilde{\tilde{A}}}_{123456}+
\tilde{\tilde{\tilde{T}}}_{125678}
 - \tilde{\tilde{\tilde{T}}}_{234578}+ \tilde{\tilde{\tilde{V}}}_{123456} \,, \nonumber\\
\tilde{\Phi}_7^{(\cal A)} &=& \tilde{\tilde{\tilde{T}}}_{234578} \,,
\nonumber \\
\tilde{\Phi}_8^{(\cal A)} &=& \tilde{\tilde{\tilde{A}}}_{123456}-2
\tilde{\tilde{\tilde{T}}}_{125678}+\tilde{\tilde{\tilde{T}}}_{234578}-\tilde{\tilde{\tilde{V}}}_{123456}
\,, \nonumber \\ \tilde{\Phi}_{9}^{(\cal A)} &=& -2
\tilde{\tilde{A}}_{34}-2 \tilde{\tilde{A}}_{123}-2
\tilde{\tilde{P}}_{21}-2 \tilde{\tilde{S}}_{12}+2
\tilde{\tilde{T}}_{127}-2
\tilde{\tilde{T}}_{158}+\tilde{\tilde{T}}_{234578}-2
\tilde{\tilde{V}}_{43}+2 \tilde{\tilde{V}}_{123} \,, \nonumber \\
\tilde{\Phi}_{10}^{(\cal A)} &=& \tilde{A}_{123}-\tilde{T}_{123}-\tilde{T}_{127}-\tilde{V}_{123} \,, \nonumber
\end{eqnarray}
\begin{eqnarray}
\tilde{\Phi}_{11}^{(\cal A)} &=& 2 \tilde{\tilde{A}}_{34}+2
\tilde{\tilde{P}}_{21}+2 \tilde{\tilde{S}}_{12}+2
\tilde{\tilde{T}}_{123}+2
\tilde{\tilde{T}}_{158}-\tilde{\tilde{T}}_{234578}+2 \tilde{\tilde{V}}_{43} \,, \nonumber\\
\tilde{\Phi}_{12}^{(\cal A)} &=& \tilde{\tilde{\tilde{T}}}_{125678}-\tilde{\tilde{\tilde{T}}}_{234578} \,, \nonumber\\
\tilde{\Phi}_{13}^{(\cal A)} &=& - \tilde{\tilde{\tilde{A}}}_{123456}+2
\tilde{\tilde{\tilde{T}}}_{125678}+
\tilde{\tilde{\tilde{V}}}_{123456}\,, \nonumber \\
\tilde{\Phi}_{14}^{(\cal A)} &=& \tilde{T}_1^M+
\tilde{\tilde{\tilde{A}}}_{123456}-
\tilde{\tilde{\tilde{T}}}_{125678}-
\tilde{\tilde{\tilde{V}}}_{123456} \,. \nonumber
\end{eqnarray}

\end{itemize}

\subsection{axial-vector transition current}
The invariant amplitudes ${\bar{F}^{(l)}_j}((p-q)^2,q^2)$ corresponding to the axial-vector transition current are obtained from the vector-current expressions by reversing the sign of the heavy-quark mass $m_b$. In addition, sign changes occur for the specific coefficient functions $\tilde{T}^{({\cal P})}_{2n}$,
$\tilde{T}^{({\cal P})}_{3n}$, $\tilde{T}^{({\cal
P})}_{6n}$, $\tilde{T}^{({\cal A})}_{1n}$,
$\tilde{T}^{({\cal A})}_{4n}$ and $\tilde{T}^{({\cal
A})}_{5n}$, as dictated by the Dirac structure of the corresponding invariant amplitudes.

\section{Differential decay width of $\Lambda_b \to \Lambda \ell^+\ell^-$}
\label{app:diff_decay_width}
For completeness, we collect the expressions used for the illustrative differential branching-fraction calculation. Retaining the operators relevant for the vector, axial-vector, and dipole contributions, the effective Hamiltonian is written as 
\begin{align}
    \mathcal{H}_{\rm eff}\ & \supset\ -\frac{4G_F}{\sqrt{2}}\, {V_{tb}V_{ts}^*} \frac{\alpha_{em}}{4\pi} \big[ C_{9}^{\rm eff} (\bar{s} \gamma_\mu P_L b) (\bar{\ell} \gamma^\mu \ell) + C_{10} (\bar{s} \gamma_\mu P_L b) (\bar{\ell} \gamma^\mu\gamma_5 \ell) \nonumber \\ &\qquad+ \frac{2 m_b}{q^2}\, C_7^{\rm eff}\,
(\bar{s} i\sigma_{\mu\nu} q^\nu P_R b)
(\bar{\ell}\gamma^\mu\ell) \big]\,, 
\label{eq:effH}
\end{align}
where $P_{L,R}=(1\mp\gamma_5)/2$. The hadronic matrix elements of the vector and axial--vector currents are expressed in terms of the helicity form factors $F\in\{f_+,f_\perp,g_+,g_\perp\}$ obtained from the pseudoscalar current LCSRs and parameterised according to Eq.~\eqref{eq:zparam}. The tensor form factors $h_+(q^2)$, $h_\perp(q^2)$ and their axial counterparts $\tilde h_+(q^2)$, $\tilde h_\perp(q^2)$ are not calculated in the present LCSR analysis and are instead taken from lattice-QCD
determinations of~\cite{Detmold:2016pkz}.

The following shorthand is used in the differential decay width,
\begin{align}
    \mathcal{N}=& \frac{G_F^2 \alpha_{em}^2 |V_{ts}^* V_{tb}|^2}{3\cdot 2^8 \pi^5 m_{\Lambda_b}^3}\,, \hspace{1cm} s_\pm = (m_{\Lambda_b} \pm m_\Lambda)^2 - q^2, \nonumber \\
     \hspace{1cm} \lambda\equiv & \lambda(m_{\Lambda_b}^2,m_{\Lambda}^2,q^2) = s_+ s_-  \,.
\end{align}
In the massless-lepton limit, the differential decay width is 
\begin{equation}
\frac{d\Gamma(\Lambda_b\to\Lambda\,\ell^+\ell^-)}{dq^2}
=
\mathcal N\;\sqrt{\lambda}\;q^2\;
\Big[2 H_T(q^2)+H_L(q^2)\Big]\,,
\label{eq:dgdq2_massless}
\end{equation}
and the transverse and longitudinal contributions are given by:
\begin{align}
H_T(q^2)
=& \Bigg[s_- \left(\left|C_9^{\rm eff}(q^2) f_\perp(q^2) + \frac{2 m_b (m_{\Lambda_b }+ m_{\Lambda})C_7^{\rm eff}}{q^2} h_\perp(q^2)\right|^2 + |C_{10} f_\perp(q^2)|^2\right)\nonumber \\ \quad & +s_+\left(\left|C_9^{\rm eff}(q^2) g_\perp(q^2) + \frac{2 m_b (m_{\Lambda_b}- m_{\Lambda})C_7^{\rm eff}}{q^2} \tilde{h}_\perp(q^2)\right|^2 + |C_{10} g_\perp(q^2)|^2\right)\Bigg]\,,
\nonumber \\[2mm]
H_L(q^2)
=&\frac{1}{q^2}\,
\Bigg[s_- (m_{\Lambda_b} + m_\Lambda)^2\left(\left|C_9^{\rm eff}(q^2) f_+(q^2) + \frac{2 m_b C_7^{\rm eff}}{(m_{\Lambda_b} + m_{\Lambda})} h_+(q^2)\right|^2 + |C_{10} f_+(q^2)|^2\right) \nonumber \\ &+s_+ (m_{\Lambda_b} - m_\Lambda)^2\left(\left|C_9^{\rm eff}(q^2) g_+(q^2) + \frac{2 m_b C_7^{\rm eff}}{(m_{\Lambda_b} - m_{\Lambda})} \tilde{h}_+(q^2)\right|^2 + |C_{10} g_+(q^2)|^2\right)\Bigg].
\label{eq:HL_massless}
\end{align}
 In the numerical analysis, $C_9^{\rm eff}(q^2)=C_9+Y(q^2)$ includes the perturbative four-quark contribution specified in Sec.~\ref{subsec:diffBR}. Genuine nonlocal hadronic matrix elements are not included in Eqs.~\eqref{eq:effH}--\eqref{eq:HL_massless}.
\bibliographystyle{JHEP}
\bibliography{references}

\end{document}